\documentclass[aps,superscriptaddress,twocolumn]{revtex4}

\usepackage{amsmath}
\usepackage{graphicx}
\usepackage{color}
\usepackage{subfigure}
\usepackage{bm}

\begin{document}

\title{Scattering resonances in slow NH${}_3$-He collisions}
\date{\today}

\author{Koos B. Gubbels}
\email{K.Gubbels@science.ru.nl}
\affiliation{Fritz-Haber-Institut der Max-Planck-Gesellschaft, Faradayweg 4-6, D-14195 Berlin,\\ Germany}
\affiliation{
Radboud University Nijmegen, Institute for Molecules and Materials,
Heyendaalseweg 135, 6525 AJ Nijmegen, \\
The Netherlands}

\author{Sebastiaan Y. T. van de Meerakker}
\affiliation{
Radboud University Nijmegen, Institute for Molecules and Materials,
Heyendaalseweg 135, 6525 AJ Nijmegen, \\
The Netherlands}
\affiliation{Fritz-Haber-Institut der Max-Planck-Gesellschaft, Faradayweg 4-6, D-14195 Berlin,\\ Germany}

\author{Gerrit C. Groenenboom}
\affiliation{
Radboud University Nijmegen, Institute for Molecules and Materials,
Heyendaalseweg 135, 6525 AJ Nijmegen, \\
The Netherlands}

\author{Gerard Meijer}
\affiliation{Fritz-Haber-Institut der Max-Planck-Gesellschaft, Faradayweg 4-6, D-14195 Berlin,\\ Germany}

\author{Ad van der Avoird}
\email{A.vanderAvoird@theochem.ru.nl}
\affiliation{
Radboud University Nijmegen, Institute for Molecules and Materials,
Heyendaalseweg 135, 6525 AJ Nijmegen, \\
The Netherlands}

\begin{abstract}

We theoretically study slow collisions of NH$_3$ molecules
with He atoms, where we focus in particular on the observation of
scattering resonances. We calculate state-to-state integral and
differential cross sections for collision energies ranging from
10${}^{-4}$ cm$^{-1}$ to 130 cm$^{-1}$, using fully converged quantum
close-coupling calculations. To describe the interaction between the
NH${}_3$ molecules and the He atoms, we present a four-dimensional
potential energy surface, based on an accurate fit of 4180 {\it ab
initio} points. Prior to collision, we consider the ammonia molecules to
be in their antisymmetric umbrella state with angular momentum $j=1$ and
projection $k=1$, which is a suitable state for Stark deceleration. We
find pronounced shape and Feshbach resonances, especially for inelastic
collisions into the symmetric umbrella state with $j=k=1$. We analyze
the observed resonant structures in detail by looking at scattering
wavefunctions, phase shifts, and lifetimes. Finally, we discuss the
prospects for observing the predicted scattering resonances in future
crossed molecular beam experiments with a Stark-decelerated NH$_3$ beam.

\end{abstract}

\maketitle

\section{Introduction}

The tremendous experimental progress in performing scattering
experiments has evolved to the point where it is nowadays possible to
study collisions between particles in the laboratory over an energy
range of about 25 orders of magnitude. The collisions of highest energy
are produced by modern charged-particle accelerators reaching the TeV
range, while the collisions of lowest energy are studied in ultracold
atomic quantum gases going all the way down to the nK regime. In both
types of collision experiments scattering resonances play an important
role. In high-energy collisions, a resonance in the cross section caused
by the formation of an intermediate bound state is a direct way to
detect previously unseen particles. In ultracold atomic scattering, the
energy of an intermediate bound molecular state to be formed during the
collision can sometimes be accurately tuned by applying an external
magnetic field. As a result, the scattering length of low-energy
$s$-wave collisions gets under full experimental control, giving rise to
a unique quantum many-body environment with a completely tunable
interaction parameter \cite{inguscio:08,stoof:09}.

Because molecules are typically harder to manipulate than atoms and
charged particles, the observations of resonances in molecular beam
scattering have been limited to a few rare cases
\cite{schutte:75a,toennies:79,skodje:00a,skodje:00b,qiu:06,Dong:Science327:1501}.
However, in recent years rapid progress has been made in performing
high-precision cold molecular scattering experiments due to the
application of the Stark deceleration technique to the study of
molecular collisions \cite{gilijamse:06}. A Stark decelerator operates
according to the same principles as a linear charged-particle
accelerator, where the dipolar or Stark force is used to decelerate
neutral polar molecules with time-varying electric fields
\cite{bethlem:99}. With the Stark decelerator it is possible
to generate almost perfectly quantum-state selected molecular beams with
a computer-controlled final velocity and a small longitudinal velocity
spread. By applying this technique to the scattering of the OH radical
with rare gas atoms, such as Xe \cite{gilijamse:06}, Ar
\cite{scharfenberg:10}, and He \cite{kirste:10}, the threshold behavior
for inelastic scattering into the first excited rotational levels of OH
could be accurately determined. Excellent agreement was found with cross
sections obtained from close-coupling calculations using {\it ab initio}
potential energy surfaces (PESs)
\cite{gilijamse:06,scharfenberg:10,scharfenberg:11}. In the same way,
also cold inelastic collisions of OH radicals with D${}_2$ molecules
were studied experimentally \cite{kirste:10}.

In this article, we study in detail cold collisions between NH${}_3$
molecules and He atoms. The ammonia-He system is a van der Waals
complex, and in general the (quasi-)bound states of such complexes are
sensitive to the interaction potential. As a result, high-resolution
spectroscopy on van der Waals complexes has been an important tool for
increasing our understanding of intermolecular forces
\cite{avoird:94,wormer:00a}. High-precision scattering experiments are a
very promising additional tool for obtaining detailed information on
potential energy surfaces. At higher scattering energies the
short-range repulsive part of the interaction is mainly probed, while
at very low collision energies the long-range part of the potential is
dominant in determining the scattering behavior. Moreover, scattering
resonances give important information on the energy of quasi-bound
states that are sensitive to potential wells at mid-range interparticle
distances. This shows that large parts of the potential energy surfaces
can be accurately probed by cold collision experiments. Recent
scattering experiments have indeed been able to distinguish between PESs
that were only of good quality and PESs that were of excellent quality
\cite{scharfenberg:10,scharfenberg:11}. A very different experiment in
which the NH$_{3}$-He interaction plays an important role, is the
trapping of NH$_{3}$ molecules inside He nanodroplets to perform
high-resolution spectroscopy \cite{slipchenko:05}.

\begin{figure}
\begin{center}
\includegraphics[width=0.6\columnwidth]{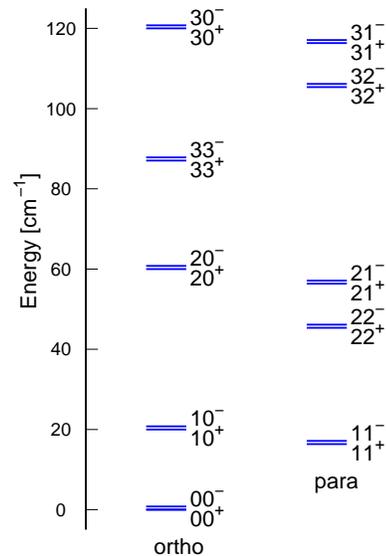}
\caption{\label{figlevels}
Energy levels $|j k \pm \rangle$ of the ammonia molecule, where $j$ is
the angular momentum of the molecule, $k$ is the projection on its
threefold symmetry axis and $+/-$ refers to the symmetric/antisymmetric
umbrella state. Throughout this article we use the $|1 1 - \rangle$
state as the initial state, so that we only consider {\it para} ammonia.
We do not take into account hyperfine interactions.
The collision energy is defined relative to the initial state.
}
\end{center}
\end{figure}

Rotational energy transfer by cold collisions is an important process in
various astrochemical environments, such as interstellar clouds and cold
exoplanetary atmospheres. Since the first identification of NH$_{3}$
molecules in the interstellar medium \cite{cheung:68}, ammonia has been
detected in several gas-phase astrochemical spectra. The rate
coefficients of NH$_{3}$-He scattering are an important
ingredient for a numerical modelling of astrochemical environments. This
is one of the reasons why NH$_{3}$-He collisions have been studied
experimentally \cite{oka:68, meyer:86, seelemann:88, schleipen:91,
meyer:95} and theoretically
\cite{green:76,billing:85,chen:97,wang:03,machin:05,yang:08} by several
groups. The most recent scattering calculations have been performed with
the potential energy surface of Hodges and Wheatley \cite{hodges:01}.
However, in order to get agreement with experimentally determined virial
coefficients, this potential had to be scaled by a rather large factor
\cite{wang:03}. The same potential has also been used to theoretically
study low-energy NH$_{3}$-He collisions, where strong scattering
resonances were observed for various initial and final states of the
NH$_{3}$ molecule \cite{yang:08}. Unfortunately, the initial state that
is most suitable for Stark deceleration was not considered. This is
namely the state $|j k \pm \rangle = |1 1 - \rangle$, where $j$ is the
angular momentum of the ammonia molecule, $k$ is the projection on its
threefold symmetry axis and $+/-$ refers to its symmetric/antisymmetric
umbrella inversion tunneling state. For the energy level diagram of the
ammonia molecule,
see Fig.~\ref{figlevels}. Moreover, in the study of Ref.~\cite{yang:08}
ammonia was treated as a rigid molecule, implying that the umbrella
inversion motion of the NH$_{3}$ molecule was not considered.

In this article we study all possible elastic and inelastic scattering
processes at low collision energies, using $| 11 - \rangle$ as an
initial state of the {\it para} NH$_{3}$ molecule. We show that
particularly the inversion inelastic scattering to the $| 11 + \rangle$
state gives rise to pronounced resonant structures that are promising to
be observed experimentally in crossed beam experiments. We start by
introducing the theoretical framework for studying the atom-molecule
collisions. After this, we present a new NH${}_3$-He potential using the
most recent developments in electronic structure calculations. We
describe the numerical methods to fit the potential, after which we
present the calculations of the integral and differential cross
sections. In both cross sections we find rapid variations as a function
of energy, which are clear signs of resonant behavior. To determine the
origin of these resonances we perform bound state calculations as well
as reconstructions of the full scattering wavefunctions. The phase
shifts and the lifetimes are also determined near resonance. Finally,
we comment on the prospects of observing these scattering resonances in
the NH${}_3$-He system in the near future.

\section{Theory}

To theoretically study the low-energy scattering of NH${}_3$
molecules with He atoms we briefly introduce three coordinate
frames that are used in the calculations \cite{bladel:91}. These
coordinate frames are discussed in more detail in the Appendix. The
first frame is an orthonormal, right-handed space-fixed (`sf')
laboratory frame located at the center of mass $\mathbf{Q}$ of the
dimer. The coordinate $R$ is the length of the vector ${\bf R}$ that
points from the center of mass $\mathbf{X}$ of the NH${}_3$ monomer to
the He atom, while $\theta^{\rm sf}$ is the zenith angle of the vector
$\mathbf{R}$ and $\phi^{\rm sf}$ is the azimuth angle in the space-fixed
frame. The second frame is an orthonormal, right-handed body-fixed
(`bf') dimer frame, also centered at the center of mass of the dimer. As
explained in the Appendix, this frame is obtained by a rotation that
aligns its $z$ axis with the vector ${\bf R}$. The third frame is an
orthonormal, right-handed monomer-fixed (`mf') frame centered at the
center-of-mass of the NH${}_3$ molecule, whose $z$ axis is aligned with
the symmetry axis of the ammonia molecule. This monomer frame is
obtained from the space-fixed frame by rotating over the three Euler
angles $\zeta^{\rm sf} = (\alpha^{\rm sf},\beta^{\rm sf},\gamma^{\rm
sf})$. Here, $\alpha^{\rm sf}$ and $\beta^{\rm sf}$ are the azimuth and
zenith angles of the ammonia $C_3$ symmetry axis in the space-fixed
frame, while $\gamma^{\rm sf}$ describes the rotation of the NH${}_3$
molecule about this axis. The Euler angles of the monomer frame can also
be given with respect to the body-fixed frame, and are then denoted by
$\zeta^{\rm bf} = (\alpha^{\rm bf},\beta^{\rm bf},\gamma^{\rm bf})$.
Finally, we introduce the umbrella or inversion angle $\rho$ of the
ammonia molecule, which is the angle between the $z$ axis of the monomer
frame and a vector pointing from the N atom to one of the H atoms, so
that $\rho = \pi/2$ corresponds to a planar NH${}_3$ geometry.

The Hamiltonian of the NH${}_3$-He system can now be written as \cite{bladel:92a}
\begin{eqnarray}\label{eq:ham}
 \hat{H} &=& \hat{H}_{\rm mol} - \frac{1}{2 \mu R} \frac{\partial^2}{\partial R^2}R + \frac{1}{2 \mu R^2}\left[ \hat{J}^2 + \hat{j}^2 - 2 \hat{\mathbf{j}}\cdot \hat{\mathbf{J}} \right] \nonumber \\
&& + V_{\rm int}(R, \beta^{\rm bf}, \gamma^{\rm bf}, \rho),
\end{eqnarray}
where throughout the article we set $\hbar = 1$, $\hat{H}_{\rm mol}$ is
the Hamiltonian of the NH${}_3$ molecule, $\mu$ is the reduced mass of
the atom-molecule complex, $\hat{\mathbf{j}}$ is the angular momentum
operator of the NH${}_3$ monomer with respect to the body-fixed frame,
$\hat{\mathbf{J}}$ is the total angular momentum operator also with
respect to the body-fixed frame, and $V_{\rm int}$ is the interaction energy.
We consider the interaction potential
to depend on four coordinates, which implies that we assume the N-H bond
length to be fixed and NH${}_3$ to keep its threefold symmetry. The
Hamiltonian of the NH${}_3$ molecule includes the monomer's rotation, as
well as the kinetic and potential energy of its umbrella motion
\cite{bladel:92a}, namely
\begin{eqnarray}
 \hat{H}_{\rm mol} &=& \sum_{\lambda=x,y,x} \frac{\hat{j}^2_{\lambda}}{2I_{\lambda\lambda}(\rho)} -\frac{1}{2\sqrt{g(\rho})} \frac{\partial}{\partial \rho} \frac{\sqrt{g(\rho})}{I_{\rho\rho}(\rho)} \frac{\partial}{\partial\rho}\nonumber\\
&&+V_{\rm umb}(\rho),\label{eq:hammon}
\end{eqnarray}
where $I_{xx}(\rho)$, $I_{yy}(\rho)$ and $I_{zz}(\rho)$ are the moments of inertia of the threefold symmetric ammonia molecule with respect to the monomer frame axes, see e.g. Ref. \cite{bladel:92a}, while $I_{\rho\rho}(\rho) = 3 m_{\rm H}r_0^2 (\cos^2 \rho + \eta \sin^2 \rho)$ with $ m_{\rm H}$ the hydrogen mass, $r_0 = 1.9099 a_0$ the fixed N-H bond length \cite{huang:08}, $\eta=m_{\rm N}/(3 m_{\rm H} + m_{\rm N})$ and $ m_{\rm N}$ the nitrogen mass. Moreover, we have that $g(\rho) = I_{xx}I_{yy}I_{zz}I_{\rho\rho}$, while the potential energy for the umbrella motion $V_{\rm umb}(\rho)$ leads to a double well potential that we model by
\begin{eqnarray}
 V_{\rm umb}(\rho) = \frac{k_{\rho}}{2}\left(\rho - \frac{\pi}{2}\right)^2+ a_{\rho} \exp \left[ - b_{\rho} \left(\rho - \frac{\pi}{2}\right)^2 \right]
\end{eqnarray}
with the parameters $k_{\rho}=90 \, 651$ cm${}^{-1}$rad${}^{-2}$, $a_{\rho} = 23 \, 229$ cm${}^{-1}$ and $b_{\rho}= 3.1846$ rad${}^{-2}$. The resulting double well potential gives rise to umbrella vibration levels of which two levels have an energy below that of the barrier at the planar ammonia geometry \cite{bladel:92a}. Moreover, each of these two vibration levels splits into a pair of energy levels due to tunneling. The parameters of the umbrella potential are chosen such that the experimental energy splitting between the two tunnel states in the vibrational ground state, as well as the experimental splittings \cite{townes:75} between the ground state and the two tunnel states of the first vibrationally excited level are accurately reproduced.

To treat the Schr\"{o}dinger equation in body-fixed coordinates, we expand the scattering wavefunction in the following coupled-channel basis
\begin{equation}
\Psi^{\rm bf} (R)= \frac{1}{R} \sum_{\mathbf{n}} | \mathbf{n}\rangle \chi_{\mathbf{n}}(R),
\end{equation}
where the radial dependence of the wavefunction is given by $\chi_{\mathbf{n}}(R)$, while the body-fixed angular basis set
\begin{eqnarray}\label{eq:bfbas}
| \mathbf{n} \rangle &\equiv&| j, k, K, J, M_J, v^{\pm} \rangle = \left[\frac{(2 j +1) (2 J + 1)}{32 \pi^3}\right]^{1/2} \nonumber \\
&&\times \phi^{\pm}_v(\rho)D^{(j)*}_{K k}(\zeta^{\rm bf})D^{(J)*}_{M_J K}(\phi^{\rm sf},\theta^{\rm sf},0)
\end{eqnarray}
is used to treat the angular part of the Hamiltonian.
Here, $D^{(j)*}_{m m'}(\zeta)=e^{i m \alpha}d^{(j)}_{mm'}(\beta)e^{i m' \gamma}$ with $d^{(j)}_{mm'}(\beta)$ the well-known Wigner $d$-functions, $k$ is the projection quantum number of the monomer angular momentum with eigenvalue $j$ on the monomer $z$ axis, $K$ is the projection quantum number of both the monomer angular momentum and the total angular momentum with eigenvalue $J$ on the body-fixed dimer $z$ axis, $M_J$ is the projection of the total angular momentum on the space-fixed $z$ axis, $v$ is the umbrella vibration quantum number, and the superscript $+/-$ refers to the even/odd umbrella tunneling function.

As a result, our task is to solve the following second-order matrix differential equation
\begin{eqnarray}\label{eq:schr}
- \frac{\partial^2 \chi_{\mathbf{n}'}(R)}{\partial R^2} = \sum_{\mathbf{n}} \langle \mathbf{n}' | \hat{W} | \mathbf{n} \rangle \chi_{\mathbf{n}}(R),
\end{eqnarray}
where we introduced the operator $\hat{W}=2\mu(E -\hat{H} + \hat{K})$
with the kinetic energy operator $\hat{K}$ given by the second term on
the right-hand side of Eq.~(\ref{eq:ham}). We note that $J$ and $M_J$
are good quantum numbers, and that the operator $\hat{W}$ is diagonal
in $J$ and independent of $M_J$. Furthermore, the monomer part of the
Hamiltonian, $\hat{H}_{\rm mol}$, is also diagonal in the angular basis
set. The complexity of the matrix $ \langle \boldsymbol{\mathbf{n}'} |
\hat{W} | \mathbf{n}\rangle$ can be further reduced by considering the
symmetry properties of the NH$_{3}$-He complex. Because the Hamiltonian
commutes with permutations of the three hydrogen atoms in NH${}_3$ and
the operator for inversion in space $\hat{E}^*$, it is useful to adapt
the basis states such that they transform as the irreducible
representations of the corresponding molecular symmetry group
$D_{3h}({\rm M})$ in the notation of Bunker and Jensen \cite{bunker:98}.
The adapted basis states of different symmetry cannot be mixed by the
Hamiltonian. The precise procedure for this adaptation is described in
the Appendix. Moreover, in Ref.~\cite{avoird:94} several useful
relations can be found for determining the matrix elements of the
$\hat{W}$ operator in the angular basis.

In order to fully solve Eq.~(\ref{eq:schr}), the wavefunctions must
satisfy the appropriate scattering boundary conditions
\cite{johnson:73}. These boundary conditions are directly formulated in
a space-fixed frame. The exact solution of the space-fixed
Schr\"{o}dinger equation at larger separations $R$, i.e., when the
interaction energy has approached zero, is a linear combination of
the proper spherical Bessel functions. These Bessel functions
are labelled by the space-fixed end-over-end rotational quantum number
$L$, which has become a good quantum number at such large separations.
Therefore, the matching of the propagated wavefunction from
Eq.~(\ref{eq:schr}) to spherical Bessel functions can be performed at
distances where the centrifugal energy, set by $L$ and decaying as
$R^{-2}$, is still large, as long as the interaction energy, decaying in
our case as $R^{-6}$, has become negligibly small. To perform the
matching, it is necessary to transform between the body-fixed and the
space-fixed basis sets. The latter basis set is for the present case
given by
\begin{eqnarray}\label{eq:sfbas}
&&| j, k, L, J, M_J, v^{\pm} \rangle = \left[\frac{(2 j +1) (2 L + 1)}{32 \pi^3}\right]^{1/2} \phi^{\pm}_v(\rho) \nonumber \\
&&\times \sum _{m_j,M_L} D^{(j)*}_{m_j k}(\zeta^{\rm sf})C^{L}_{M_L}(\theta^{\rm sf},\phi^{\rm sf}) \langle j m_j; L M_L | J M_J \rangle,\quad ~
\end{eqnarray}
with $C^{L}_{M_L}(\theta^{\rm sf},\phi^{\rm sf})$ the Racah-normalized spherical harmonics and $M_L$ the projection of the end-over-end angular momentum on the space-fixed $z$ axis, $m_j$ the projection of the monomer angular momentum on the space-fixed $z$ axis, and $\langle j_1 m_1; j_2 m_2 | j_3 m_3 \rangle$ a Clebsch-Gordan coefficient. The transformation between the body-fixed and the space-fixed basis then becomes \cite{avoird:94}
\begin{eqnarray}
&&| j, k, L, J, M_J, v^{\pm} \rangle \\
&&= \sum_{K} | j, k, K, J, M_J, v^{\pm} \rangle \left(\frac{2 L + 1}{2J+1}\right)^{1/2} \langle j K; L 0 | J K \rangle.\nonumber
\end{eqnarray}
To end our discussion of the matching procedure, we mention the various
possible open and closed channels following from the Hamiltonian of
Eq.~(\ref{eq:hammon}). The vibration-tunneling states of the
umbrella motion are determined by calculating the eigenstates of the
Hamiltonian formed by the last two terms of Eq.~(\ref{eq:hammon}). Only
the lowest four eigenstates, labeled $\phi_{v}^{\pm}(\rho)$ with
vibrational quantum numbers $v=0$ and $v=1$, are kept. With these four
states as a basis for the umbrella motion, we turn to the first term of
the Hamiltonian of Eq. (\ref{eq:hammon}). As a result, the rotational
constants $1/2I_{\lambda \lambda}(\rho)$ become $4\times4$ matrices, but
the $+$ and $-$ states are not mixed. The eigenstates that result from
diagonalization of the full monomer Hamiltonian in a basis containing all
rotational states with $j \le 6$ and the four umbrella states, determine
the open and closed channels. We label the open channels by $|j k \pm
\rangle$; the vibrational quantum number $v$ is omitted from this
notation because all vibrational states with $v>0$ are closed for the
energy range in which we are interested. We consider the ammonia
molecules to be prepared in the $| 1 1 - \rangle$ state, so that the
lower lying $| 1 1 + \rangle$ state is open for all collision
energies. Increasing the collision energy beyond the energy of excited
monomer states opens up the corresponding channels, and inelastic
scattering into these states occurs if it is allowed by symmetry. The
matching procedure of the wavefunctions to the boundary conditions for
scattering at large distance $R$ ultimately leads to an expression for
the scattering matrix. This $S$ matrix is subsequently directly related
to the differential and integral cross sections for the elastic and
inelastic channels, which can be compared with the outcome of collision
experiments \cite{child:74}.

\section{The NH${}_3$-He potential}

Before we can apply the above described formalism to solve the
scattering problem, we need to determine the NH${}_3$-He interaction
potential. To this end {\it ab initio} calculations were performed with
{\sc molpro} \cite{molpro:09}, using the supermolecule approach with the
counterpoise procedure of Boys and Bernardi \cite{boys:70}. We
considered the interaction energy to be dependent on four coordinates,
namely $R$, $\beta^{\rm bf}$, $\gamma^{\rm bf}$ and $\rho$.

The grid for the {\it ab initio} calculations consisted of 4180 points.
For $R$, in total 19 points were used. In the short and intermediate
range, i.e., for $4 a_0 \le R \le 10 a_0$, we used an equidistant grid of
13 points with a separation of $0.5 a_0$, while in the long range, that
is for $R > 10 a_0$, we used an approximately logarithmic grid
consisting of the points $12 a_0$, $14.4 a_0$, $17.3 a_0$, $20.8 a_0$,
$25 a_0$, and $30 a_0$. For $\beta^{\rm bf}$, we used a Gauss-Legendre
grid consisting of 11 points for $0 \le \beta^{\rm bf} \le \pi$, while
for $\gamma^{\rm bf}$, we used an equidistant Gauss-Chebyshev grid
consisting of the points $\pi/24$, $3\pi/24$, $5\pi/24$, and $7\pi/24$.
Finally, for the grid in $\rho$ we used an equidistant grid of 5 points,
where the middle point was given by the value $\rho_{3}=0.6226 \pi$,
while the distance between the points was given by $\Delta\rho=( 2
\rho_{3}-\pi)/5$.

The calculations in the long range were performed with the
coupled-cluster method taking into account single and double excitations
and a perturbative treatment of triple excitations [CCSD(T)], using the
augmented correlation-consistent polarized valence quadruple-zeta (AVQZ)
basis set. For the short range we used the explicitly correlated
CCSD(T)-F12 method \cite{adler:07} to account more efficiently for the
strong effect of electron correlations in this regime. The CCSD(T)-F12
method was found to yield accurate results with the smaller
triple-zeta basis set (AVTZ), as illustrated by Table \ref{tabf12}. In
this table the interaction energies are shown for different NH${}_3$-He
geometries in the short, intermediate and long range with both the
CCSD(T) and the CCSD(T)-F12 method using different basis sets. Also the
effect of using midbond functions \cite{koch:98} is included in this table,
where the midbond orbitals were located along the vector connecting the
center of mass of the ammonia molecule with the helium atom at a
distance of $(r_0+R)/2$ from the ammonia center of mass. Particularly in
the short and intermediate range these midbond functions improve the
interaction energies, so that they were used in the calculations with
F12. From Table \ref{tabf12}, we see that in the short range the
CCSD(T)-F12 method with an AVTZ basis set including midbond functions
performs better than the CCSD(T) method with an AV5Z basis set, although
the latter calculation is much more expensive due to the large basis
set.

The main reason why we did not use the F12 method in the long range is
that we found that the implementation of this method in {\sc molpro}
\cite{molpro:09} gives rise to an incorrect $1/R$ behavior in the very
long range, rather than the correct $1/R^6$ behavior for the system
under consideration. This behavior is caused by the fitting of the
electron density distributions, which unfortunately does not result in
exactly
charge neutral monomers. Although the artifical residual charges can be
reduced by introducing a larger electron density fitting basis set,
the $1/R$ behavior will eventually always dominate the correct
$1/R^6$ behavior. Hence, we decided to use the CCSD(T) method
without F12 for the long range.
We used the AVQZ basis set, and we may conclude from Table
\ref{tabf12} that this basis set indeed gives rise to accurate
interaction energies in the long range. In order to switch smoothly
between the results of the two methods, we used the switching function
$s(R)$
\begin{eqnarray}
s(R)=\left\{
\begin{array}{l l}
0 & {\rm if } \quad R\le a \\
1 & {\rm if } \quad R\ge b \\
 \frac{1}{2}+\frac{1}{4}\sin\frac{\pi x}{2}(3-\sin^2\frac{\pi x}{2}) &  {\rm otherwise}
 \end{array}
 \right.
\end{eqnarray}
with $x=(2R -b-a)/(b-a)$, $a =10 a_0$ and $b=13 a_0$. The function is chosen such that the first three derivatives of $s$ at $R=a$ and $R=b$ are zero. We thus calculated the interaction energies for the angular geometries at distance $R=12 a_0$ with both methods, where the calculated value with F12 was given a weight of $1-s(12a_0)$, while the value without F12 was given a weight $s(12a_0)$.

\begin{table}[t]
\caption{\label{tabf12} Comparison of the interaction energy between the CCSD(T) method (abbreviated as CC) and the CCSD(T)-F12 method (abbreviated as F12) for different basis sets and different geometries as calculated with {\sc molpro} \cite{molpro:09}. We used the augmented triple zeta (AVTZ), quadruple zeta (AVQZ) and quintuple zeta (AV5Z) basis sets. We also studied the effect of midbond functions \cite{koch:98}, which are indicated in the Table by the $+$ sign, when they are added to the basis set. For the short-range geometry, indicated by ${\bf x}_{\rm s}$ in the Table, we used $R = 4.5 a_0$, $\beta^{\rm sf}= 0$, $\gamma^{\rm sf}= 0$ and $\rho =14 \pi/24$. For the mid-range geometry, indicated by ${\bf x}_{\rm m}$, we used $R = 7 a_0$, $\beta^{\rm sf}= \pi/2$, $\gamma^{\rm sf}= \pi/6$ and $\rho =15 \pi/24$. For the long-range geometry, indicated by ${\bf x}_{\rm l}$, we used $R = 15 a_0$, $\beta^{\rm sf}= \pi$, $\gamma^{\rm sf}= \pi/3$ and $\rho = 16 \pi/24$. The interaction energies are given in
cm$^{-1}$.  }
\begin{tabular}{ l  c  c  c  c  c  }
 & AVTZ  & AVTZ$+$  & AVQZ  & AVQZ$+$ & AV5Z  \\
\hline
CC (${\bf x}_{\rm s}$) &   1446.2& 1414.2  & 1407.4  & 1397.3  & 1394.7  \\
F12 (${\bf x}_{\rm s}$)&   1393.5& 1386.0  & 1389.8  & 1386.8  & 1386.1  \\
CC (${\bf x}_{\rm m}$) &$-$21.716&$-$23.529&$-$22.733&$-$23.521&$-$23.179\\
F12 (${\bf x}_{\rm m}$)&$-$23.329&$-$23.679&$-$23.237&$-$23.510&$-$23.381\\
CC (${\bf x}_{\rm l}$) &$-$0.2506&$-$0.2583&$-$0.2521&$-$0.2538&$-$0.2526\\
F12 (${\bf x}_{\rm l}$)&$-$0.2622&$-$0.2637&$-$0.2548&$-$0.2554&$-$0.2524\\
\end{tabular}
\end{table}

\begin{figure}
\begin{center}
\includegraphics[width=0.65\columnwidth]{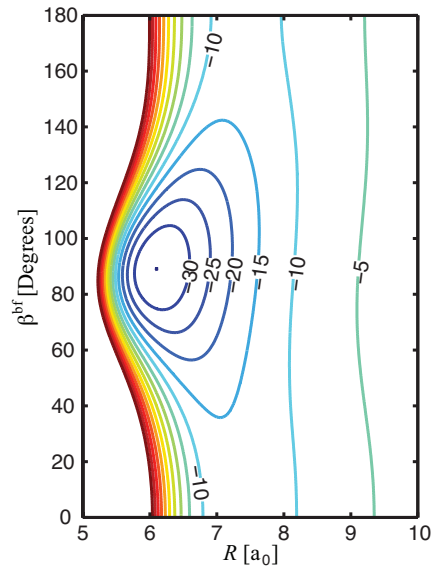}
\caption{\label{figpot} Contour plot of the  NH${}_3$-He interaction potential as a function of $R$ and $\beta^{\rm bf}$ for $\gamma^{\rm bf} = 0$ and the equilibrium umbrella angle $\rho_{\rm e} = 112.15 {}^{\circ}$. The energies of the contours are given in cm$^{-1}$. }
\end{center}
\end{figure}

To obtain an analytic representation of the interaction potential
between the NH${}_3$ molecules and the He atoms we first perform an
expansion in tesseral spherical harmonics, namely
\begin{equation}
\label{vaniso}
V_{\rm int}(R, \beta^{\rm bf}, \gamma^{\rm bf}, \rho) =\sum_{l,m} (-1)^m v_{lm}(R,\rho) S_{lm}( \beta^{\rm bf}, \gamma^{\rm bf}),
\end{equation}
where due to the symmetry of the dimer only terms with $m = 0,3,6,...$
are present. Because we have 11 grid points in $\beta^{\rm bf}$, the
summation over $l$ is from $0 \le l \le 10$. The summation over $m$ is
from 0 to the largest multiple of 3 that is smaller than or equal to the
corresponding $l$ value. On all grid points $R_i$ and $\rho_j$ we
determine the coefficients of the angular expansion $v_{lm}(R_i,\rho_j)$
by means of a quadrature on the {\it ab initio} grid with the
appropriate Gauss-Legendre and Gauss-Chebyshev weights \cite{avoird:94}.
For the resulting expansion coefficients $v_{lm}$, we distinguish
between the short-range and the long-range behavior, so that
$v_{lm}(R,\rho)= v^{\rm sr}_{lm}(R,\rho)+v^{\rm lr}_{lm}(R,\rho)$. Both
in the short range and the long range the dependence of the coefficients
$v_{lm}$ on $\rho$ is represented by a polynomial expansion in
$(\rho-\pi/2)^p$, where $p$ ranges from $0$ to $9$. If $l+m$ is even,
then the polynomial expansion only contains even powers in $p$, while if
$l+m$ is odd only odd powers are present. In the long range, we expanded
the potential in powers of $R^{-n}$, resulting in
\begin{equation}\label{eqcoeflr}
v^{\rm lr}_{lm}(R, \rho)  =\sum_{n,p} c_{lmpn} f_{n}(a R) \left(\rho-\frac{\pi}{2}\right)^p R^{-n}.
\end{equation}
where the inverse powers of $R$ that are involved depend on $l$. It can
be shown \cite{avoird:80} that for $l = 0,2$ the expansion starts with
$n_i =6$, while for $l=1,3$ it starts with $n_i=7$. For $l \ge 4$, it
starts with $n_i = l + 4$. We used the analytic long range expansion of
Eq. (\ref{eqcoeflr}) only for $l \le 5$, and for each $l$ we took the
leading term $R^{-n_i}$ and the next-to-leading term $R^{-n_i-2}$ into
account. The Tang-Toennies damping function
\begin{equation}
f_n(x)=1-\left(\sum_{i=0}^n  \frac{x^i}{i!} \right)e^{-x}.
\end{equation}
was included to avoid the singular behavior of the long-range terms in
the short range \cite{tang:84}. For the value of $a$ in
Eq.~(\ref{eqcoeflr}) we used the isotropic exponent in the short-range,
or, to be more precise, $a=
\ln[v_{00}(R_1,\rho_3)/v_{00}(R_2,\rho_3)]/\Delta R =2.088 a_0^{-1}$
with $R_1$ and $R_2$ the first two points of the $R$ grid and $\Delta
R=R_2-R_1$. The expansion coefficients $c_{lmpn}$ were obtained from
$v_{lm}(R_i,\rho_j)$ by performing a weighted least-squares fit using
the last three points of the $R$ grid and all points of the $\rho$ grid.
The three $R$ points were weighted for each $l$ by $R_i^{n_i}$, with
$n_i$ the leading power of the long range decay for the considered $l$.

To describe the short and intermediate range of the potential the same expansion was employed in $\beta^{\rm bf}, \gamma^{\rm bf}$ and $\rho$ as for the long range, but the behavior in $R$ was treated differently. The following procedure was used. First, for all the grid geometries, the corresponding value of the analytic long range potential was subtracted from the {\it ab initio} values. Next, after performing the expansion in tesseral harmonics and powers of $(\rho-\pi/2)$,  the behavior of the resulting coefficients $v_{lmp}(R)$ was interpolated with a reproducing kernel Hilbert space (RKHS) method \cite{ho:96}. The smoothness parameter of the RKHS interpolation was set to 2, while the RKHS parameter $m$, which determines the power with which the interpolation function decays, was chosen to depend on $l$. For $l \le 5$, the parameter was set to $m = n_i+1$. Then the RKHS function decays as $R^{-n_i-2}$, which is faster for each $l$ than the leading term in the analytic fit of Eq. (\ref{eqcoeflr}).
However, for $l>5$, no analytic long range fit was done, and we used $m=n_i -1$. As a result, the corresponding RKHS functions decayed for each $l$ as $R^{-n_i}$ with $n_i = l+4$, which is the correct leading long range behavior for $l > 5$ \cite{avoird:80}.

We have compared the fitted potential with the {\it ab initio} values on the full grid to test the accuracy of the fit in the angles $\beta^{\rm bf}, \gamma^{\rm bf}$ and $\rho$. The quality of the fit in $R$ cannot be tested in this way, because the RKHS procedure goes by construction precisely through the points to be fitted. We calculated the  RMS (root mean square) error for each grid distance $R_i$ and divided by the mean {\it ab initio} interaction energy at that distance, giving for the relative RMS error $\xi(R_i)$ that
\begin{eqnarray}
\xi(R_i) &=&  \frac{\sqrt{\frac{1}{n}\sum_{j,k,l}[\Delta V(R_i,\beta_j^{\rm bf}, \gamma_k^{\rm bf},\rho_l)]^2}}{
\left|\frac{1}{n}\sum_{j,k,l} V_{\rm int}^{\rm abi}(R_i,\beta_j^{\rm bf}, \gamma_k^{\rm bf},\rho_l)\right| } \,100 \%,\quad
\end{eqnarray}
where $\Delta V(R_i,\beta_j^{\rm bf}, \gamma_k^{\rm bf},\rho_l)= V^{\rm fit}_{\rm int}(R_i,\beta_j^{\rm bf}, \gamma_k^{\rm bf},\rho_l)-V^{\rm abi}_{\rm int}(R_i,\beta_j^{\rm bf}, \gamma_k^{\rm bf},\rho_l)$, and the summations are over all $n=220$ angular grid points. For our
potential fit we found that the relative error $\xi(R_i)$ is less than
0.05 \% for all $R_i$, so the fits in the angular coordinates are
excellent. To also test the fit in $R$, we calculated {\it ab initio}
interaction energies for an additional 495 points, that were chosen to
lie about halfway between the grid points used for the fit. The
relative RMS error of the values calculated from the fit compared to the
new {\it ab initio}
values depended quite strongly on $R$, where the largest error was found
to occur in the short range. Namely, for the test points at $R=4.3 a_0$
we found with the use of 45 different angular points a relative RMS
error of $3.5\%$, while for all other $R$ values we obtained a relative
error of about $0.5\%$ or less. An important reason for this behavior is
that we use a RKHS fit for the short range, which behaves as a power
law, while the true behavior of the potential is exponential. The
fitting procedure could thus have been further improved using an
exponential form. However, we note that already the present fitting
error is rather small. Moreover, in the present paper we use the
potential to describe cold scattering with collision energies of
maximally $130$ cm$^{-1}$, so that the extreme short-range behavior of
the potential is not being probed.

\begin{figure}
\begin{center}
\includegraphics[width=0.9\columnwidth]{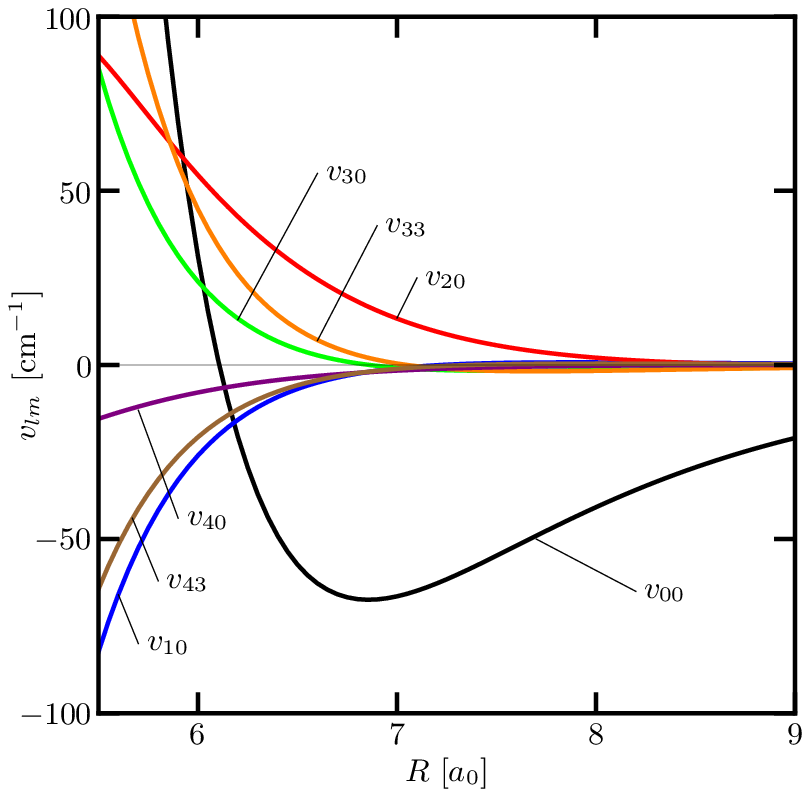}
\caption{\label{figvlm} Coefficients $v_{lm}(R,\rho_e)$ of the NH${}_3$-He interaction energy as a function of center-of-mass distance $R$, evaluated at the equilibrium umbrella angle $\rho_e = 112.15 {}^{\circ}$. The isotropic $v_{00}(R,\rho_e)$ coefficient is largest. The $v_{10}$, $v_{20}$, $v_{30}$, $v_{33}$, $v_{40}$ and $v_{43}$ coeffients are shown as well.}
\end{center}
\end{figure}

In Fig.~\ref{figpot}, we show a contour plot of the NH${}_3$-He
interaction potential for $\gamma^{\rm bf} = 0$ and the equilibrium umbrella angle $\rho_e =
112.15 {}^{\circ}$. For this value of $\gamma^{\rm bf}$ and $\rho$, the
minimum of the potential is given by $V_{\rm min }=-35.08$ cm$^{-1}$ for
$R=6.095$ $a_0$ and $\beta^{\rm bf} = 89.0{}^{\circ}$. This may be
compared to the potential of Hodges {\it et al.} \cite{hodges:01}, where
the minimum of the potential for $\gamma^{\rm bf} = 0$ and $\rho_e =
112.15 {}^{\circ}$ is given by $V_{\rm min }=-33.46$ cm$^{-1}$ for
$R=6.133$ $a_0$ and $\beta^{\rm bf} = 88.75{}^{\circ}$. Although this
difference in the well depth is not very large, we have found that the
consequences for low-energy scattering can still be quite
substantial, as we will discuss in Section \ref{secresults}. Finally, we
have for the leading isotropic coefficient, defined as $C_6 =
-v_{00}(R,\rho_3) R^6 $ for large $R$, that in atomic units $C_6 = 39.6$
$E_h \cdot a_0^6$. The relative importance of the various $v_{l
m}(R,\rho_e)$ expansion coefficients is shown in Fig.~\ref{figvlm}. The
potential is available in {\sc fortran} 77 on {\sc epaps} \cite{epaps:pot}.

\section{Computational aspects}

Having discussed the formalism and the potential, we now turn to the
numerical procedures that we used in order to obtain converged cross
sections that can be compared with future cold-collision experiments. To
numerically determine the four lowest lying vibration-inversion levels
$\phi^{\pm}_{v}(\rho)$ of the Hamiltonian of Eq.~(\ref{eq:hammon}) for
$j = 0$, we used the discrete variable representation based on
sinc-functions (sinc-DVR) \cite{groenenboom:93}. The resulting
eigenfunctions were used to determine the matrix elements $\langle \phi
'|1/2I_{\lambda\lambda}| \phi \rangle$ with $\phi =\phi_{v}^{\pm}(\rho)
$ by numerical integration. For the propagation in solving
Eq.~(\ref{eq:schr}), the renormalized Numerov algorithm was used,
starting at $4 a_0$ and ending at $50 a_0$, using an equidistant grid
with 273 points. The renormalized Numerov method also allows for a
complete reconstruction of the scattering wavefunctions.

The angular basis set contained all monomer states with $j \le 6$, where
we checked that the inclusion of more monomer levels resulted only in
deviations of maximally 1 \% for the calculated cross sections. The
maximal value for the total angular momentum $J$ that we used depended
on the collision energy. For collision energies $E \le 10$ cm${}^{-1}$,
we included all angular basis states with $J \le 10$, while for $10< E
\le 50$ cm${}^{-1}$, we included all basis states with $J \le 20$, and
for $50< E \le 130$ cm${}^{-1}$, we included all states with $J \le 30$.
The convergence of the cross sections with respect to the total angular
momentum $J$ is slowest for the elastic cross section. The inelastic
cross sections are converged at considerably lower values of $J$ than
reported here.

In order to check our results and gain additional insight, we also implemented a commonly applied model to treat the ammonia  umbrella motion in scattering calculations \cite{davis:78,green:80}. In this model, no vibrationally excited umbrella states are taken into account and the ground-state umbrella tunneling states are approximated as an even and odd combination of the two rigid equilibrium structures. These two states are thus written as $| \pm \rangle = [f(\rho-\rho_e)\pm f(\pi-\rho+\rho_e)]/2^{1/2}$, where $f(x)$ is a function localized around $x=0$. More precisely, the two-state model amounts to approximating the matrix elements of the potential by $\langle \pm | v_{lm}(R,\rho) | \pm\rangle = v_{lm}(R,\rho_e)$ for $l+m$ even, and $\langle \pm | v_{lm}(R,\rho) | \mp\rangle = v_{lm}(R,\rho_e)$ for $l+m$ odd. For the rotational constants we use the experimentally determined values $A_{xx} =
A_{yy} = 9.9402$ cm${}^{-1}$, and  $A_{zz}=6.3044$ cm${}^{-1}$ in the model. Furthermore, we include the experimental ground state splitting of $0.79$ cm$^{-1}$ \cite{townes:75} between the two tunneling states in the scattering calculations. This simple model has been implemented in the scattering program {\sc molscat} \cite{molscat:94}. We have used {\sc molscat} to double-check the results that we obtained from our own scattering program. The model was previously found to result in good agreement with more elaborate treatments of the umbrella motion for scattering at higher collision energies \cite{sanden:93}. In this article we also want to test the accuracy of the model for cold collisions, and in particular for the calculation of scattering resonances.

\section{Results} \label{secresults}

\begin{figure}
\begin{center}
\includegraphics[width=1.0\columnwidth]{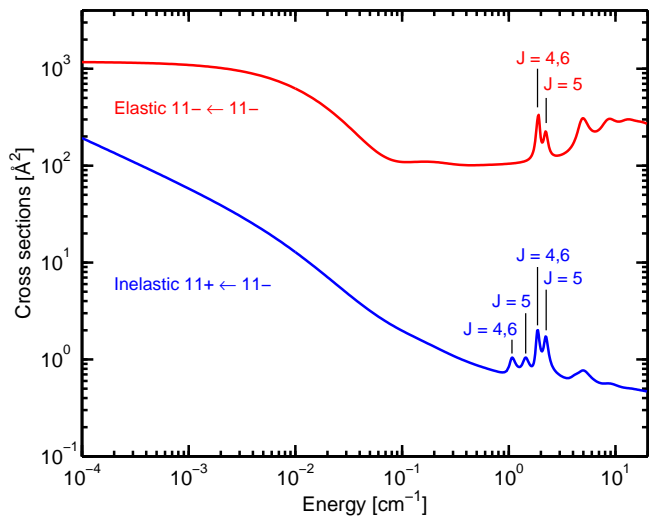}
\caption{\label{figcross_log} Integral cross sections for NH${}_3$-He scattering as a function of collision energy.  The initial state of the NH${}_{3}$ molecule is the $|11-\rangle$ state. At low collision energies only elastic scattering (upper red curve) and inelastic scattering into the lower lying $|11+\rangle$ state (lower blue curve) can occur.}
\end{center}
\end{figure}

In Fig.~\ref{figcross_log}, we show the integral cross sections for the
scattering of NH${}_3$ molecules with He atoms for collision energies
ranging from 10${}^{-4}$ cm${}^{-1}$ to 20 cm${}^{-1}$.
Initially, we only have elastic scattering and inelastic scattering into
the $| 1 1 + \rangle$ state, which lies $0.79$ cm${}^{-1}$ lower in
energy. Figure~\ref{figcross_log} was made using the previously described
elaborate treatment of the umbrella motion, however, with the use of the
model treatment almost exactly the same results were obtained. We
observe in the first place that, in agreement with the Wigner threshold
laws \cite{wigner:48}, the elastic cross section becomes constant for very small
collision energies, while the inelastic cross section decreases with $E$
as $1/\sqrt{E}$. Going more into the details of the figure, we observe
two shape resonances closely together in the elastic channel at
collision energies of 1.86 and 2.22 cm${}^{-1}$. In bound state
calculations with the NH$_3$-He complex enclosed in a box of variable
size we found continuum levels with nearly the same energies that are
practically independent of the box size, so we may conclude that these
peaks in the scattering cross section indeed correspond to shape or
orbiting resonances caused by quasi-bound states. Such quasi-bound
states may occur either in the incoming or in the outgoing scattering
channel; for the specific case of elastic scattering these are the same.
Looking at the dominant contributions to the cross section, the first
peak was found to be mainly caused by quasi-bound states with total
angular momenta $J=4$ and $J=6$, while the second peak was mainly caused
by a quasi-bound state with total angular momentum $J=5$. In both cases
they corresponded to an end-over-end angular momentum of
$L=5$. Looking in the same energy range at the inelastic scattering into
the $| 1 1 + \rangle$ state, we observe not only two similar peaks
at the same collision energies, but also two additional peaks at 1.08
and 1.44 cm${}^{-1}$. These two additional shape resonances can be
readily understood by noting that for inelastic scattering the resonant
quasi-bound state can occur either in the incoming channel or in the
outgoing channel, where the latter channel is about $0.8$ cm$^{-1}$
lower in energy. This is indeed precisely the energy with which the two
additional peaks in the inelastic channel are shifted to the left in
Fig.~\ref{figcross_log}.

\begin{figure}
\begin{center}
\includegraphics[width=1.0\columnwidth]{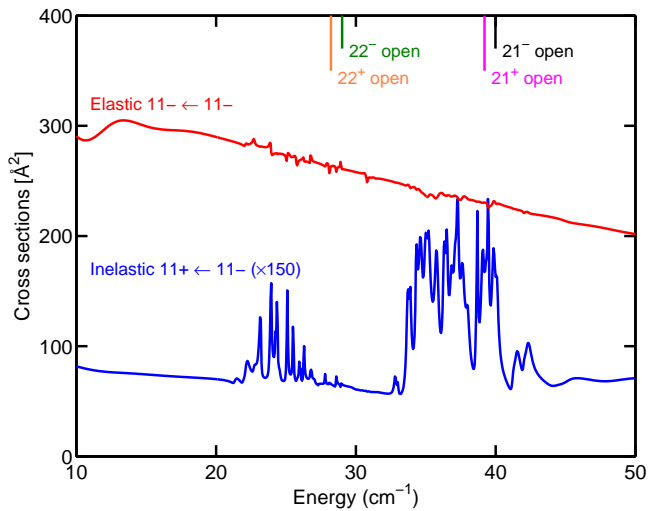}
\caption{\label{figcross1}
Elastic (upper red curve) and inversion inelastic (lower blue curve)
integral cross sections as a function of collision energy for
NH${}_3$-He scattering. The inelastic cross
section is scaled with a factor of 150, so that the actual
inelastic cross section is 150 times smaller than shown in the figure.
At higher collision energies the $|2 2 \pm\rangle$ and the $|2 1
\pm\rangle$ channels open. As a result, Feshbach resonances are
observed, which are most pronounced for the inelastic scattering into
the $|11+\rangle$ state.}
\end{center}
\end{figure}

For completeness, we note that we studied this collision energy range
also with the potential of Hodges and Wheatley \cite{hodges:01}.
Although the difference in the well depth between the two potentials at
the equilibrium umbrella angle was only about 5\%, we still found large
differences in the resonant structures at very low energies. For
example, using the Hodges and Wheatley potential \cite{hodges:01}, we
observed two very strong shape resonances at collision energies of 0.03
and 0.45 cm${}^{-1}$ induced by quasi-bound states with total angular
momenta $J=3$ and $J=4$ and end-over-end angular momentum $L=4$.
However, because our own potential is deeper, we find that these
quasi-bound states have become true bound states with energies below the
scattering continuum, so that they cannot cause shape resonances
anymore. As a result, the first shape resonances we find with our
potential are induced by quasi-bound states with total angular momenta
$J=4,5$ and $6$ and $L=5$, as shown in Fig.~\ref{figcross_log}. This
point also clearly shows that scattering resonances at low energy can be
very sensitive to the precise shape of the potential energy surface,
which means that accurate scattering experiments can be used to probe
very precisely our knowledge of intermolecular interactions.

\begin{figure}
\begin{center}
\includegraphics[width=1.0\columnwidth]{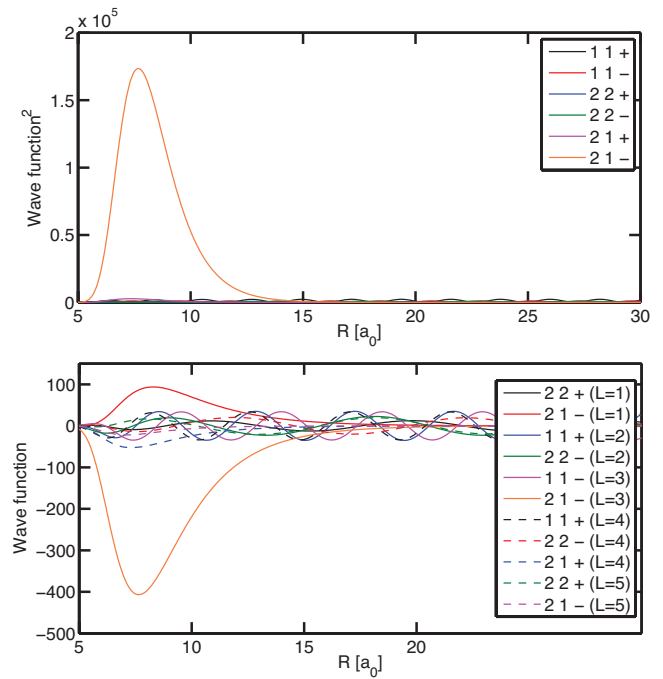}
\caption{\label{figwf1}
(a) Contributions of various channels and partial waves to the
scattering wavefunction at a collision energy of 37.28 cm${}^{-1}$,
where a Feshbach resonance for inelastic scattering into the $| 1 1 +
\rangle$ state occurs. We used a total angular momentum of $J=3$ and
considered $E''$ symmetry. As a result, the initial state $|11-\rangle$
asymptotically corresponds to $L=3$, while there are eight open partial
waves in outgoing channels.
(b) Contributions of various channels to the square of the wavefunction.
In the inner region, where the collision takes place, a large amplitude
in the asymptotically closed $|21-\rangle$ channel is observed. }
\end{center}
\end{figure}

In Fig.~\ref{figcross1}, we again show the integral cross sections for
elastic scattering and inelastic scattering into the $| 1 1 + \rangle$
state, but now considering collision energies from 10 to 50
cm${}^{-1}$. Note that the inelastic cross is actually
150 times smaller than shown in the figure. As can be seen
from Fig.~\ref{figlevels}, at a collision energy of $28.33$
cm$^{-1}$ it becomes energetically possible to excite the ammonia
molecule from its $| 1 1 - \rangle$ state to its $| 2 2 + \rangle$
state, and a new scattering channel opens. At 29.12, 39.33 and 40.12
cm$^{-1}$, the $| 2 2 - \rangle$, $| 2 1 + \rangle$ and $| 2 1 -
\rangle$ channels open, respectively, as also indicated in
Fig.~\ref{figcross1}. The opening of the new channels is seen to have a
profound effect on the inelastic cross sections to the $| 1 1 + \rangle$
state. Namely, before these new channels open a bunch of Feshbach
resonances is observed. These resonant structures are called Feshbach
resonances because they are caused by a molecular level that
is different from the incoming and the outgoing channel. In
Fig.~\ref{figcross1} we see that especially the Feshbach resonances
induced by the $| 2 1 \pm \rangle$ levels at collision energies around
40 cm$^{-1}$ are strong, giving rise to almost a factor of 3 increase
compared to the background inelastic cross section. These resonances
seem to be particularly suited to observe in a collision experiment. We
come back to this point more elaborately in Section \ref{sec:disc}.

\begin{figure}
\begin{center}
\includegraphics[width=1.0\columnwidth]{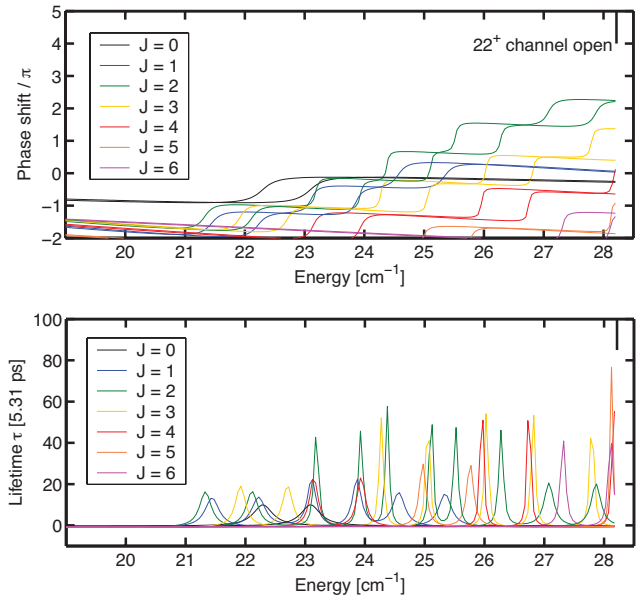}
\caption{\label{figphases}
Top panel: Phase shift sum as a function of collision energy for various
total angular momenta $J$. Both $E'$ and $E''$ symmetries are
considered, so that we have two curves for each $J$. Bottom panel: the
corresponding lifetimes as a function of the collision energy. The
lifetimes are obtained from the derivative of the phase shifts with
respect to the collision energy (in cm$^{-1}$); they are given in units
of 5.3088 ps. }
\end{center}
\end{figure}

To understand the Feshbach resonances in more detail, we have studied
the scattering wavefunctions. In Fig.~\ref{figwf1} contributions to the
scattering wavefunction are shown at a collision energy of 37.28
cm${}^{-1}$. At this collision energy, there is a Feshbach resonance for
inelastic scattering into the $| 1 1 + \rangle$ state, somewhat below
the energy at which the $| 2 1 - \rangle$ channel opens. In
Fig.~\ref{figwf1}(a), we show contributions of different open and closed
channels to the scattering wavefunction. For this particular figure, we
considered a total angular momentum of $J=3$ and symmetry $E''$ (see the
Appendix). This means that for the incoming channel, i.e., the
$|11-\rangle$ state, asymptotically only the partial wave with $L = 3$
contributed. For the four open outgoing channels, namely $|11\pm\rangle$
and $|22\pm \rangle$, in total eight open partial waves are possible for
the considered $J$ and $E''$ symmetry. In the inner region also
contributions corresponding to asymptotically closed channels can gain
amplitude, when they are coupled to the considered incoming state and
outgoing state by the interaction potential. In Fig.~\ref{figwf1}(b), we
show for each channel the resulting contributions to the square of the
wavefunction. From Figs.~\ref{figwf1}(a) and (b) we clearly see that in
particular the closed $|21-\rangle$ channel has a very strong amplitude
in the collision region, which shows that this state is responsible for
the strong Feshbach resonance observed at this collision energy.

A different way to study the Feshbach resonances is by looking at phase
shifts in the scattering wavefunction. These phase shifts can be
obtained from the eigenvalues of the scattering matrix \cite{child:74,ashton:83}. In
Fig.~\ref{figphases}, we show in the top panel the sums of the phase
shifts in all open channels for various total angular momenta $J$. Since
we consider both symmetries $E'$ and $E''$, we have two curves for each
$J$. From scattering theory it follows that when a resonance occurs, the
phase shift sum rapidly increases by $\pi$ \cite{ashton:83} as a function
of energy. In the top panel of Fig.~\ref{figphases}, we indeed see this
happening many times at the collision energies where resonances are
found in the elastic and inelastic cross sections. The derivatives of
the phase shifts with respect to the energy give the lifetime of the
collision complex \cite{child:74}. These lifetimes are shown in the
lower panel of Fig.~\ref{figphases}. This figure shows that at the
collision energies where resonances occur, we indeed have long-lived
intermediate quasi-bound states.

\begin{figure}
\begin{center}
\includegraphics[width=1.0\columnwidth]{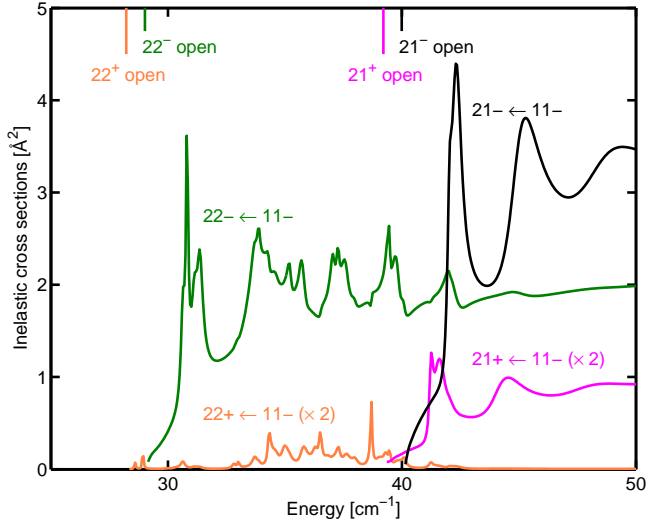}
\caption{\label{figcross2}
Inelastic integral cross sections for scattering into the $|2 2
\pm\rangle$ channels and the $|2 1 \pm\rangle$ channels as a function of
collision energy. The inelastic cross sections for the $| 2 2 + \rangle$
and the $| 2 1 + \rangle$ channels are scaled with a factor of 2.
After the various channels open, shape resonances are observed.}
\end{center}
\end{figure}

In Fig.~\ref{figcross2}, we show the integral cross sections for
inelastic scattering into the $| 2 1 \pm \rangle$ and the $| 2 2 \pm
\rangle$ states, for a collision energy ranging from the energies at
which these channels open, up to 50 cm${}^{-1}$. Note that the inelastic
cross sections to the $| 2 2 + \rangle$ and $| 2 1 + \rangle$ states are
scaled with a factor of 2. Immediately after each channel opens, we
see strong resonant features, which are shape resonances, caused by
quasi-bound states in the outgoing channel. In the $| 2 2 \pm \rangle$
channels we also find Feshbach resonances due to quasi-bound states of
$| 2 1 \pm \rangle$ character.

In the energy range from 10 to 50 cm${}^{-1}$, we also studied the
scattering cross sections using the previously described model treatment
of the NH${}_{3}$ umbrella motion. We found that the model calculations
have the tendency to somewhat overestimate the strength of certain resonance
peaks compared to the elaborate treatment of the umbrella motion.
Studying this effect in more detail, we found that the differences are
mainly due to the approximation of the nonzero potential matrix elements for the two tunneling states as $v_{lm}(R,\rho_e)$, rather than due to the neglect of
the higher lying $\phi_1^{\pm}(\rho)$ states. Namely, by calculating the
cross sections with the elaborate treatment and taking only the lowest
two umbrella functions $\phi_0^{\pm}(\rho)$ into account we obtained
cross sections that were nearly equal to the elaborate treatment with
four umbrella functions, while they gave rise to the same differences
with the model treatment. However, we note that in general the model
treatment performed very satisfactory in describing the resonance
structures. All resonant peaks found with the elaborate treatment were
also found with the model treatment, and typically the strength of the
scattering resonances differed by less than 10\%. Because the precise
strength and location of the resonances are very sensitive to the
the potential, we conclude that the use of the model
treatment is useful in studying scattering resonances, especially in
cases when the elaborate treatment is computationally too expensive.

\begin{figure}
\begin{center}
\includegraphics[width=1.0\columnwidth]{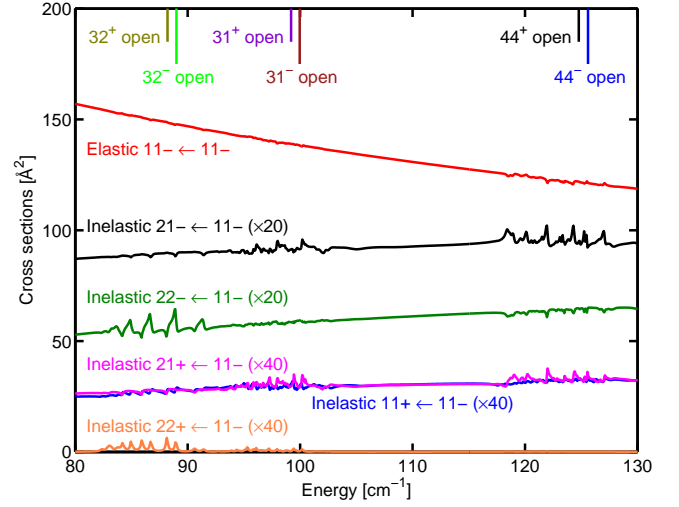}
\caption{\label{figcross3}
Elastic and inelastic integral cross sections for scattering into the
$|1 1 +\rangle$, $|2 2 \pm\rangle$, $|2 1 \pm\rangle$ states as a
function of collision energy. The
inelastic cross sections for the $| 2 2 - \rangle$ and $| 2 1 -
\rangle$ channels are scaled with a factor of 20, and for the $|1 1
+\rangle$, $|2 2 +\rangle$ and $|2 1 +\rangle$ channels with a factor of
40. At higher collision energies various $|3 k \pm\rangle$ and $|4 k
\pm\rangle$ channels for the {\it para} ammonia molecules open. As a
result, small Feshbach resonances are observed.}
\end{center}
\end{figure}

Looking again at Fig.~\ref{figcross2}, we note that there are
significant differences in the magnitudes of the inelastic cross
sections for the various collision channels. For example, the transition
to the $| 2 2 - \rangle$ state is seen to be much stronger than the
transition to the $| 2 2 + \rangle$ state, and the same holds for the
transition to the $| 2 1 - \rangle$ state compared to the $| 2 1 +
\rangle$ state. The relative magnitude of the integral cross sections
for the elastic and inelastic scattering channels can even be more
clearly observed in Fig.~\ref{figcross3}. In this figure, we show the
integral cross sections for scattering into the $| 1 1 \pm \rangle$, $|
2 1 \pm \rangle$ and $| 2 2 \pm \rangle$ states, for collision energies
ranging from 80 cm${}^{-1}$ to 130 cm${}^{-1}$. Notice the scaling of
the inelastic cross sections indicated in the figure. To explain the
relative strengths of the transitions shown, we note that the scattering
from the $| 1 1 - \rangle$ channel into different $| j k\pm \rangle$
channels is caused by different anisotropic terms in the interaction
potential with coefficients $v_{lm}(R,\rho)$, cf. Eq.~(\ref{vaniso}).
For example, in order to change the umbrella state of the ammonia
molecule (i.e., going from the odd $-$ state to the even $+$ state) we
need terms in the potential for which $l+m $ is odd, so that also the
corresponding coefficient $v_{lm}(R,\rho)$ is odd in $\rho$. The various
potential energy coefficients are plotted as a function of $R$ at the
equilibrium umbrella angle $\rho_e$ in Fig.~\ref{figvlm}.

\begin{figure}
\begin{center}
\includegraphics[width=1.0\columnwidth]{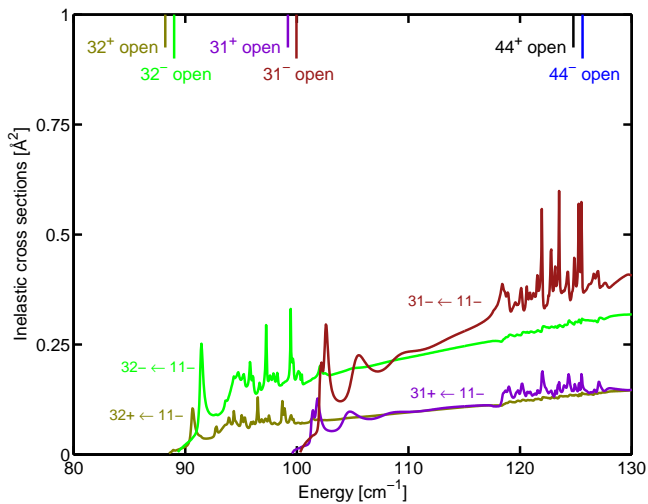}
\caption{\label{figcross4}
Inelastic integral cross sections for scattering into the $|3 2
\pm\rangle$ and the $|3 1 \pm\rangle$ states as a function of collision
energy. At higher collision energies the $|4 4 \pm\rangle$ channels for the {\it
para} ammonia molecules open. As a result, Feshbach resonances are
observed.}
\end{center}
\end{figure}

From this figure, we see that the isotropic coefficient
$v_{00}(R,\rho_e)$ is by far the largest coefficient of all. This
coefficient can only couple the initial $|1 1 - \rangle$ state to itself
(see for example Refs.~\cite{avoird:94,millan:95}), causing a large
elastic scattering cross section. For a transition to a different
umbrella state, or to a state with different $j$, we need potential
energy coefficients with $l \ge 1$. Since $v_{10}$ and $v_{30}$ are odd
in $\rho$, they cause transitions from the $|1 1 - \rangle$ state to the
$|2 1 + \rangle$ and the $|1 1 + \rangle$ state for example. From
Fig.~\ref{figcross3}, we see that the inelastic cross sections to these
two states are indeed approximately equally large. The $v_{20}$ term causes transitions to the $|2 1 - \rangle$ state, and because
this expansion coefficient is relatively large, the corresponding cross
section is large as well. Finally, in order to change $k$ in the
collision, we need potential terms with $m \ne 0$, of which the first
two are the $v_{33}$ and the $v_{43}$ coefficients. The $v_{33}$
coefficient causes $- \rightarrow -$ transitions and the $v_{43}$
coefficient causes $- \rightarrow +$ transitions. From Fig.~\ref{figvlm}
we see that the $v_{33}$ coefficient is rather large, explaining the
large cross sections to the $|2 2 - \rangle$ state, while the $v_{43}$
coefficient is small, explaining the small cross sections to the $|2 2 +
\rangle$ state.

\begin{figure*}
\begin{center}
\includegraphics[width=1.4\columnwidth]{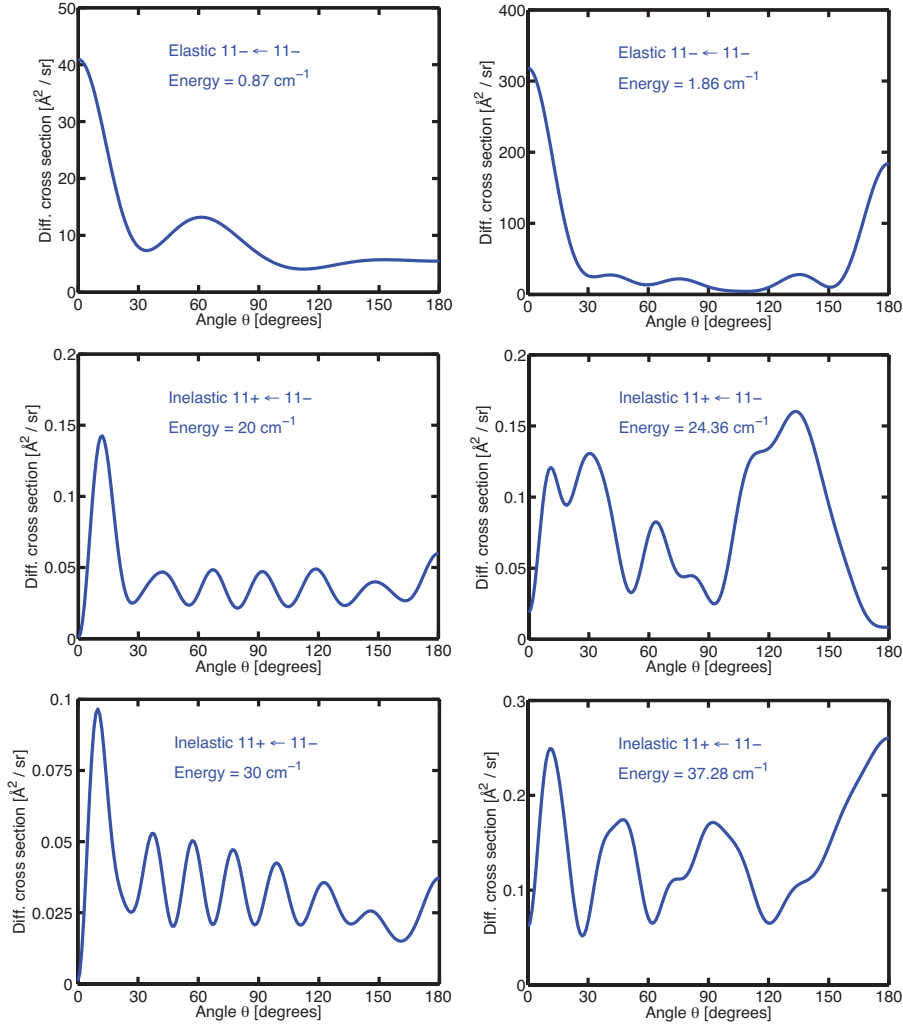}
\caption{\label{figdcs} Differential cross sections as a function of the zenith angle $\theta$ for collisions between NH${}_{3}$ and He at various collision energies. Upper two panels: differential cross sections for elastic scattering at a collision energy of 0.87  cm${}^{-1}$ (left) and 1.86 cm${}^{-1}$ (right). In the first case there is no resonance, the second case corresponds to a shape resonance. Middle two panels: differential cross sections for inelastic scattering into the $| 1 1 + \rangle$ state at a collision energy of 20  cm${}^{-1}$ (left panel, no resonance) and 24.36 cm${}^{-1}$ (right panel, Feshbach resonance). Lower two panels: the same but for a collision energy of 30  cm${}^{-1}$ (left panel, no resonance) and 37.28 cm${}^{-1}$ (right panel, Feshbach resonance).}
\end{center}
\end{figure*}

In Fig.~\ref{figcross4}, we show the integral cross sections for
inelastic scattering into the $| 3 2 \pm \rangle$ and the $| 3 1 \pm
\rangle$ states, for a collision energy ranging from the energies at
which these channels open, up to 130 cm${}^{-1}$. These small
cross sections will be harder to observe experimentally. However, if
these channels can be detected, they allow for the observation of
pronounced shape resonances at higher collision energies. The cross
sections for scattering into the $| 3 2 \pm \rangle$ states also give
rise to Feshbach resonances with quasi-bound states of $| 3 1 \pm
\rangle$ character.
Even stronger Feshbach resonances at higher collision energies between
about 120 and 125 cm${}^{-1}$ are found in the cross sections for
scattering into the $| 3 1 - \rangle$ state. These Feshbach resonances
are caused by the opening of the $| 4 4 \pm \rangle$ channels at a
collision energy of about 125 cm${}^{-1}$.

Finally, we also studied differential cross sections, where we looked in
particular at the behavior of the differential cross sections as a
function of energy close to resonance. In general, we found that the
differential cross sections can change rapidly and dramatically close to
resonance. This behavior is illustrated by Fig.~\ref{figdcs}. Here, we
see in the upper two panels the differential cross sections for elastic
scattering at collision energies of 0.87 and 1.86 cm${}^{-1}$. For the
first of these energies there is no resonance, while for the second
there is a shape resonance. For the off-resonance case we find that,
apart from diffraction oscillations, there is only a forward scattering
peak. On resonance there is also a strong backward peak. The lower four
panels show the differential cross sections for inelastic scattering to
the $| 1 1 + \rangle$ state at collision energies of 20, 24.36, 30 and
37.28 cm${}^{-1}$. At 20 and 30 cm${}^{-1}$, which are shown in the two
lower plots on the left, there is no resonance and the differential
cross sections look rather similar to the upper left one, giving
predominantly rise to forward scattering. At 24.36 and 37.28
cm${}^{-1}$, which are shown in the two lower plots on the right, there
is a Feshbach resonance present, and as a result the differential cross
sections look very differently, giving again rise to significant
backscattering. In general, the precise structure of the differential
cross section depends on various aspects such as the lifetime and the
rotational state of the intermediate collision complex. As a result, it
is expected that the differential cross sections show clear changes near
a resonance, but the precise way in which they change is hard to predict
and can be very different for different resonances, as is also seen in
Fig.~\ref{figdcs}.

\section{Discussion and conclusion}\label{sec:disc}

In this article, we have theoretically studied cold collisions
of NH${}_3$ molecules with He atoms, where we looked in detail at shape
and Feshbach scattering resonances. Prior to collision, we considered
the ammonia molecules to be in their antisymmetric umbrella state with
angular momentum $j=1$ and projection $k=1$, which is a suitable state
for Stark deceleration. We calculated state-to-state integral and
differential cross sections for collision energies ranging from
10${}^{-4}$ cm$^{-1}$ to 130 cm$^{-1}$, using fully converged quantum
close-coupling calculations. We treated the umbrella motion of the
ammonia molecule by solving the corresponding Hamiltonian in curvilinear
coordinates and taking the resulting first four vibration-tunneling
states exactly into account. We call this the elaborate treatment. We
also used a common model for the umbrella motion which approximates the
umbrella tunneling states as an even and odd combination of the two
possible rigid equilibrium structures for ammonia. This we call the
model treatment.

To describe the interaction between the NH${}_3$ molecules and the He
atoms accurately, we presented a new four-dimensional potential energy
surface, based on a high-quality fit of 4180 {\it ab initio} points. In
the short range we used the explicitly correlated CCSD(T)-F12 method
with an AVTZ basis set including midbond functions, and we showed that
this approach leads to excellent results in the short range. In the long
range we used the CCSD(T) method with an AVQZ basis but without F12,
since we found that the electron density fitting that accompanies the
F12 treatment does not exactly preserve charge neutrality of the
monomers and eventually leads to a dominant $1/R$ dependence of the
potential at very large $R$ values. Our potential has a well depth $D_{e} = 35.08$ cm${}^{-1}$, which is to be compared with the well depth of 33.46 cm${}^{-1}$ for the potential of Hodges and Wheatley
\cite{hodges:01}. Although this difference is not very large, we found
that small differences in the potential can have profound
consequences for the observed resonance structures at low scattering
energies.

We studied all open collision channels for {\it para} ammonia up to
$j=3$ and in all these channels we found pronounced shape resonances
right after the opening of these channels, caused by quasi-bound states
in the incoming and outgoing channels. We also found Feshbach
resonances that are particularly strong for the outgoing
$| 1 1 + \rangle$ channel at collision energies of about 25 cm${}^{-1}$
caused by intermediate $| 2 2 \pm \rangle$ states, and at collision
energies of about 35 cm${}^{-1}$ caused by intermediate $| 2 1 \pm
\rangle$ states. Due to the large cross section of these inelastic
resonances, namely more than 1 \AA${}^2$, they seem to be a good
candidate for experimental observation. Also in the $| 3 1 - \rangle$
channel at collision energies of about 120 cm${}^{-1}$
relatively strong Feshbach resonances were seen that are due to
intermediate $| 4 4 \pm \rangle$ states. We analyzed the observed resonant structures in
detail by looking at the corresponding scattering wavefunctions, phase
shifts and lifetimes. We also investigated the validity of using the
model treatment for the ammonia umbrella motion in describing
low-energy scattering resonances. We found that the model performs
qualitatively very well, but on a quantitative level some resonance
peaks are somewhat overestimated compared to the elaborate treatment.
However, considering the sensitivity of these resonances to the
interaction potential, for which even state-of-the-art {\it ab initio}
methods still lead to uncertainties on the order of a percent, the
model treatment seems adequate in treating low-energy resonant
scattering, especially in cases when the elaborate treatment becomes
computationally too expensive.

The calculated integral cross sections at low collision energies can be
measured using Stark-decelerated molecular beams. The NH$_3$ molecule,
and its isotopologue ND$_3$, are amenable to the Stark deceleration
technique, and have been employed frequently in deceleration experiments
\cite{Bethlem:PRA65:053416}. The Stark decelerator provides a beam of
ammonia molecules, state-selected in the upper inversion component of
the $j=k=1$ level, with a velocity that is tunable between standstill
and high velocities \cite{Heiner:PCCP8:2666}. In a crossed beam
experiment, the Stark decelerated ammonia molecules can be collided with
an atomic beam of helium. In an optimized geometry, the two beams
collide at a small beam intersection angle. An intersection angle of
less than $ 90^{\circ}$ reduces the attainable collision energy and
improves the collision energy resolution of the experiment
\cite{Scharfenberg:PCCP13:8448}. As shown in section \ref{secresults},
there are a number of scattering channels with pronounced shape and/or
Feshbach resonances. The most promising prospects for the experimental
observation of resonant features are found in the channels $|11+ \rangle
\leftarrow |11- \rangle$ (see Fig.~\ref{figcross1}), $|22- \rangle
\leftarrow |11- \rangle$, and $|21- \rangle \leftarrow |11- \rangle$ (see
Fig.~\ref{figcross2}).

\begin{figure}
\begin{center}
\includegraphics[width=1.0\columnwidth]{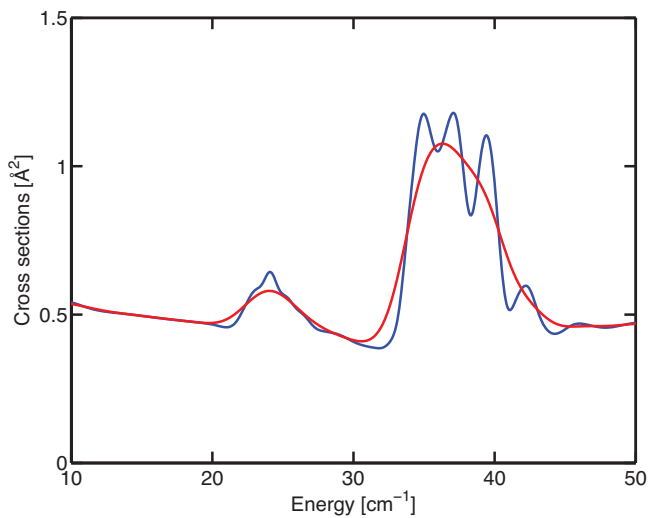}
\caption{\label{figconv}
Convoluted inversion inelastic integral cross sections as a function of
the mean collision energy for NH${}_3$-He scattering. The initial state
of the NH${}_{3}$ molecules is the $|11-\rangle$ state and the final
state is the $|11+\rangle$ state. The figure is similar to
Fig.~\ref{figcross1}, only now we have assumed a Gaussian collision
energy distribution for the colliding particles to simulate more
realistically what would be observed with present day experimental
technology. The blue curve corresponds to a full width at half maximum
(FWHM) of $1~$cm$^{-1}$, and the red curve of $3~$cm$^{-1}$.
}
\end{center}
\end{figure}

The resonant structures are found at collision energies in the 20 - 50
cm$^{-1}$ range. To simulate what would be observed in a molecular beam
scattering experiment, we convoluted the integral scattering cross
section for the $|11+\rangle \leftarrow |11-\rangle$ inversion inelastic channel with
Gaussian collision energy distributions having both a 1~cm$^{-1}$ and a
3~cm$^{-1}$ full width at half maximum (FWHM). In the considered
scattering channel bunches of Feshbach resonances are observed that are
caused by the opening of the $|22\pm\rangle$ and $|21\pm\rangle$ channels, as seen in
Fig.~\ref{figcross1}. The result of the convolutions are shown in
Fig.~\ref{figconv}. From this figure we conclude that the
details of the dense resonance structures in Fig.~\ref{figcross1} can
only be resolved when an experimental collision energy spread that is
much less than $1$ cm$^{-1}$ can be achieved. For an experimental
resolution of 3~cm$^{-1}$, however, the bunch of scattering resonances
can still be discerned from the background inelastic signal by measuring
the inelastic cross section as a function of collision energy.
Figure~\ref{figconv} shows that such a measurement would lead to
a clear enhancement of the inelastic signal by more than a factor of two
at the energies where the bunch of resonances is located.

To estimate the feasibility of obtaining collision energy resolutions in
this range with current experimental technology, we assume an
experiment in which a Stark-decelerated packet of NH$_3$
molecules collides with a conventional beam of He atoms at a beam
intersection angle of 45$^{\circ}$. We assume that the He atom beam is
produced using a cryogenic source that is maintained at a temperature of
about 50 Kelvin, resulting in a He atom velocity of 550 m/s. The
relevant range of collision energies is then obtained when the velocity
of the NH$_3$ molecules is tuned between 370 and 850 m/s. This is well
within the range of state-of-the-art Stark deceleration molecular beam
machines \cite{Scharfenberg:PRA79:023410}. We further assume velocity
spreads of 10 m/s and 55 m/s for the NH$_3$ molecules and He atoms,
respectively, and a spread in the beam intersection angle of 40 mrad due
to the divergence of both beams. These are values that can realistically be
obtained with current experimental techniques. With these parameters, we
expect an optimum in the collision energy resolution to occur at a
collision energy of 30 cm$^{-1}$, i.e., at the center of the relevant
collision energy range. This maximum accuracy amounts to a spread of 3.1
cm$^{-1}$ (FWHM), while the collision energy spread increases to
approximately 4 cm$^{-1}$ both for collision energies down to 20
cm$^{-1}$ and collision energies up to 50 cm$^{-1}$. These energy
resolutions will not yet allow for the observation of single scattering
resonances in the NH$_3$-He system, but they will certainly enable to
observe the enhancement of about a factor of two in the inelastic cross
section as a function of collision energy shown in Fig.~\ref{figconv},
revealing the combined effect of the underlying bunch of Feshbach
resonances.

An alternative and complementary approach to study scattering resonances
is to measure differential cross sections. Referring back to Fig.~\ref{figdcs}, dramatic changes in the differential cross section can
occur at collision energies where a resonance is observed. Feshbach
resonances that give rise to strong backward scattering can be detected
by measuring selectively the scattered flux in the backward direction. A
similar approach has been used recently to measure partial-wave resolved
resonances in the collision energy dependence of reactions between F
atoms and HD molecules \cite{Dong:Science327:1501}. For inelastic
scattering between NH$_3$ molecules and He atoms, differential cross
sections are measured most conveniently using the velocity map imaging
(VMI) technique \cite{eppink:97}. To experimentally resolve the angular
dependence of the differential cross sections, large recoil velocities
of the scattered molecules are advantageous. For the $|11+\rangle \leftarrow
|11-\rangle$ channel, the recoil velocity of the scattered NH$_3$ molecules in
the center of mass frame amounts to about 100~m/s at the most relevant
collision energies. This is well within the range of velocities that can
be imaged using current VMI techniques, offering interesting prospects
to study the behavior of molecular scattering resonances.

At the Fritz-Haber-Institute in Berlin, Germany, and the Radboud
University Nijmegen, The Netherlands, we have embarked on an
experimental program to study scattering resonances in both the integral
and differential cross sections using Stark-decelerated molecular beams.
It is the hope that the experimental study of these resonances will test
our theoretical understanding of molecular interactions with
unprecedented accuracy, and also will contribute to an enhancement of our ability to control the way in which molecules collide.

\appendix

\section{Coordinates and symmetry}

\begin{table*}[t]
\caption{\label{tabcoor} Transformation properties of the coordinates under symmetry operations.}
\begin{tabular}{@{\extracolsep{0.3cm}} l c c c c c c c c c }
Operation	& \multicolumn{8}{c}{Effect on angular coordinates} & \\
\hline
$\hat{E}$ & $\theta^{\rm sf}$ & $\phi^{\rm sf}$ & $\rho$ & $\alpha^{\rm sf}$ & $\beta^{\rm sf}$ & $\gamma^{\rm sf}$ & $\alpha^{\rm bf}$ & $\beta^{\rm bf}$ & $\gamma^{\rm bf}$ \\
$(123)$ & $\theta^{\rm sf}$ & $\phi^{\rm sf}$ & $\rho$ & $\alpha^{\rm sf}$ & $\beta^{\rm sf}$ & $\gamma^{\rm sf}-2\pi/3$ & $\alpha^{\rm bf}$ & $\beta^{\rm bf}$ & $\gamma^{\rm bf}-2\pi/3$  \\	
$(23)^*$ & $\pi-\theta^{\rm sf}$ & $\phi^{\rm sf} + \pi$ & $\rho$ & $\alpha^{\rm sf}+\pi$ & $\pi-\beta^{\rm sf}$  & $\pi-\gamma^{\rm sf}$ & $\pi-\alpha^{\rm bf}$ & $\beta^{\rm bf}$  & $-\gamma^{\rm bf}$  \\
$\hat{E}^*$ &  $\pi-\theta^{\rm sf}$ & $\phi^{\rm sf} + \pi$ & $\pi - \rho$ & $\alpha^{\rm sf}$ & $\beta^{\rm sf}$ & $\gamma^{\rm sf}+\pi$ & $-\alpha^{\rm bf}$ & $\pi-\beta^{\rm bf}$ & $\gamma^{\rm bf}$  \\
$(123)^*$ & $\pi-\theta^{\rm sf}$ & $\phi^{\rm sf} + \pi$  & $\pi - \rho$ & $\alpha^{\rm sf}$ & $\beta^{\rm sf}$ & $\gamma^{\rm sf}+\pi/3$ & $-\alpha^{\rm bf}$ & $\pi-\beta^{\rm bf}$ & $\gamma^{\rm bf}-2\pi/3$ \\
$(23)$ &  $\theta^{\rm sf}$ & $\phi^{\rm sf}$ & $\pi - \rho$ & $\alpha^{\rm sf} + \pi$ & $\pi-\beta^{\rm sf}$ & $-\gamma^{\rm sf}$ & $\alpha^{\rm bf} + \pi$ & $\pi-\beta^{\rm bf}$ & $-\gamma^{\rm bf}$
\end{tabular}
\end{table*}

In this Appendix we consider in more detail the various coordinates used to describe the NH${}_3$-He system and how these coordinates transform under various symmetry operations that commute with the Hamiltonian. These symmetry operations form a group generated by the permutations of the hydrogen atoms in NH${}_3$ and the operator for inversion in space $\hat{E}^*$. The location of the center-of-mass of the NH${}_3$-He dimer is given by the vector ${\bf Q}$, while the locations of the nuclei with respect to ${\bf Q}$ are given by the vectors ${\bf H}_1$, ${\bf H}_2$, ${\bf H}_3$ for the three H atoms, by the vector ${\bf N}$ for the N atom, and by ${\bf He}$ for the He atom. The center-of-mass of the ammonia molecule is given by the vector ${\bf X}$. We start by introducing a orthonormal, right-handed space-fixed (`sf') frame centered at the center-of-mass of the dimer ${\bf Q}$. We also make the convention that a superscript denotes the frame in which the coordinates of a vector are expressed. So in the space-fixed frame an arbitrary vector ${\bf P}$ has the space-fixed coordinates ${\bf P}^{\rm sf}$. Moreover, when no frame superscript is given, we do not specify the frame in which the coordinates of the vector ${\bf P}$ are evaluated. The space-fixed frame consists of three unit vectors, which are the columns of the matrix ${\bf s} = ({\bf s}_x,{\bf s}_y,{\bf s}_z)$. In our notation ${\bf s}^{\rm sf}$ is the unit matrix. Moreover, in the space-fixed frame the inversion operator $\hat{E}^*$ maps the position of any nucleus ${\bf P}^{\rm sf}$ onto the position reflected in the origin $-{\bf P}^{\rm sf}$.

Another useful frame, the dimer frame or body-fixed frame ${\bf d} = ({\bf d}_x,{\bf d}_y,{\bf d}_z)$, is obtained by performing two rotations to the space-fixed frame in order to align the ${\bf d}_z$ axis of the dimer frame along the vector ${\bf R} = {\bf He} - {\bf X} $, which points from the center-of-mass of the ammonia molecule to the helium atom. We have in space-fixed coordinates that
\begin{eqnarray}\label{eqrsf}
{\bf R}^{\rm sf}=\left(\begin{array}{c}
R \cos \phi^{\rm sf} \sin \theta^{\rm sf} \\
R \sin \phi^{\rm sf} \sin \theta^{\rm sf} \\
R \cos \theta^{\rm sf}
\end{array} \right),
\end{eqnarray}
so that $\phi^{\rm sf}$ and $\theta^{\rm sf}$ are the azimuth and zenith
angle of the vector ${\bf R}$ in the space-fixed frame. The body-fixed
frame is then defined in the following way. For any vector ${\bf P}$, we
have that ${\bf P} = {\bf s} \cdot {\bf P}^{\rm sf}={\bf d} \cdot {\bf
P}^{\rm bf}$ with ${\bf P}^{\rm sf}=\boldsymbol{\mathcal{R}}_{\rm
bf}^{\rm sf}\cdot {\bf P}^{\rm bf}$ and $\boldsymbol{\mathcal{R}}_{\rm
bf}^{\rm sf} = \boldsymbol{\mathcal{R}}_{z} (\phi^{\rm sf})
\boldsymbol{\mathcal{R}}_{y}(\theta^{\rm sf})$, where
$\boldsymbol{\mathcal{R}}_{y}(\theta^{\rm sf})$ and
$\boldsymbol{\mathcal{R}}_{z}(\phi^{\rm sf})$ are the usual rotation
matrices for rotation about the $y$ axis and the $z$ axis, respectively,
with the convention that $\boldsymbol{\mathcal{R}}_z(\phi^{\rm sf})_{12} = -\sin
\phi^{\rm sf}$ and $\boldsymbol{\mathcal{R}}_y(\theta^{\rm sf})_{31} = -\sin \theta^{\rm
sf}$. Note that as a result we find for the coordinates of the vector
${\bf R}$ in the body-fixed frame that $(\boldsymbol{\mathcal{R}}_{\rm
bf}^{\rm sf}) ^T{\bf R}^{\rm sf} = {\bf R}^{\rm bf}$ resulting in
$ R^{\rm bf}_{x}= R^{\rm bf}_{y}=0$ and $ R^{\rm bf}_{z}=R$, as was required. When we apply the $\hat{E}^*$ operator to
the complex, the coordinates ${\bf R}^{\rm sf}$ get inverted, so that
$\hat{E}^*$: $ {\bf R}^{\rm sf} \rightarrow-{\bf R}^{\rm sf}$. As a
result, the angles in Eq. (\ref{eqrsf}) are changed according to
$\phi^{\rm sf} \rightarrow \phi^{\rm sf}+\pi$ and $\theta^{\rm sf}
\rightarrow \pi-\theta^{\rm sf}$. Note that the dimer frame is invariant
under permutations of the hydrogen atoms.

\begin{table*}[t]
\caption{\label{tabbas} Effect of the symmetry operations on the angular basis functions.}
\begin{tabular}{@{\extracolsep{0.5cm}} l  r  r }
Operation $\qquad$ & Body-fixed &  Space-fixed  \\
\hline
$\hat{E}$ & $| j k K J M v^{\pm} \rangle $ & $| j k L J M v^{\pm} \rangle $ \\
$(123)$ &  $e^{2\pi i k /3} | j k K J M v^{\pm} \rangle $  & $e^{2\pi i k /3} | j k L J M v^{\pm} \rangle $ \\
$(23)^*$ & $ \pm (-1)^{J+k} | j {-k} {-K} J M v^{\pm} \rangle $ & $\pm (-1)^{j+k+L}| j {-k} L J M v^{\pm} \rangle$\\
$\hat{E}^*$ & $\pm (-1)^{J+j+k}| j k {-K} J M v^{\pm} \rangle $ &  $\pm (-1)^{L+k}| j k L J M v^{\pm} \rangle $ \\
$(123)^*$ & $ \pm (-1)^{J+j+k}e^{2\pi i k /3} | j k {-K} J M v^{\pm} \rangle $ & $\pm e^{2\pi i k /3} | j k L J M v^{\pm} \rangle $ \\
$(23)$ & $(-1)^{j} | j {-k} K J M v^{\pm} \rangle $ & $(-1)^{j} | j {-k} L J M v^{\pm} \rangle$
\end{tabular}
\end{table*}

The third useful frame is called the monomer frame and it is located at the center-of-mass of the ammonia molecule ${\bf X}$. The frame is spanned by the vectors
\begin{eqnarray}\label{eqfmon}
{\bf v}_x &=& 2{\bf H}_1 -{\bf H}_2-{\bf H}_3, \nonumber \\
{\bf v}_y &=& {\bf H}_2 -{\bf H}_3,  \\
{\bf v}_z &=& {\bf v}_x \times {\bf v}_y. \nonumber
\end{eqnarray}
Since the ammonia molecule keeps it threefold symmetry, the above frame
is orthogonal. The monomer frame can consequently be made orthonormal by
dividing the vectors in Eq. (\ref{eqfmon}) by their length, resulting in
the right-handed monomer frame denoted by ${\bf f}$. The rotation matrix
that expresses the monomer frame axes in space-fixed coordinates is
given in terms of the three Euler angles $\zeta^{\rm sf}=(\alpha^{\rm
sf},\beta^{\rm sf},\gamma^{\rm sf})$, resulting in
$\boldsymbol{\mathcal{R}}^{\rm sf}_{\rm
mf}=\boldsymbol{\mathcal{R}}_{z}(\alpha^{\rm
sf})\boldsymbol{\mathcal{R}}_{y}(\beta^{\rm
sf})\boldsymbol{\mathcal{R}}_{z}(\gamma^{\rm sf})$. When the $\hat{E}^*$
operator is applied to the complex, then both the $x$ axis and the $y$
axis of the monomer frame are reversed in the space-fixed frame, which
leaves the $z$ axis in place. As a result, the $\hat{E}^*$ operator has
the following effect on the Euler angles, $\hat{E}^*$: $\alpha^{\rm sf}
\rightarrow \alpha^{\rm sf}$, $\beta^{\rm sf} \rightarrow \beta^{\rm
sf}$, and $\gamma^{\rm sf} \rightarrow \gamma^{\rm sf} + \pi$. Another
angle that is important in our treatment of the ammonia-helium complex
is the inversion or umbrella angle $\rho$, defined as the angle between
the $z$ axis of the monomer frame and the vector pointing from the
N atom to one of the H atoms. So, for
$\rho=\pi/2$, ammonia has a planar geometry. We just showed that
$\hat{E}^*$ leaves the $z$ axis in place, while the coordinates of the
nuclei change sign. As a result, we have that $\hat{E}^*$: $\rho
\rightarrow \pi - \rho$.

When we permute the hydrogen nuclei, the space-fixed coordinates of the
monomer frame axes are interchanged. For example, when we interchange
$\vec{{\bf H}}_2$ and $\vec{{\bf H}}_3$ we find that the monomer
$y$ axis and also the $z$ axis are reversed in the space-fixed frame. As
a result, the corresponding Euler angles transform as $(23)$:
$\alpha^{\rm sf} \rightarrow \pi + \alpha^{\rm sf}$, $\beta^{\rm sf}
\rightarrow \pi - \beta^{\rm sf}$, and $\gamma^{\rm sf} \rightarrow
-\gamma^{\rm sf}$. Since the $(23)$ operation inverts the monomer $z$
axis, we also find that $(23)$: $\rho \rightarrow \pi-\rho$. In Table
\ref{tabcoor} we show the transformation properties of the various
angles that describe the ammonia-helium complex when symmetry operations
of the permution-inversion group $D_{3h}({\rm M})$ are applied to the
complex. These transformation properties are useful in determining the
transformation properties of the angular basis functions.

The monomer frame can also be obtained by a rotation from the body-fixed
dimer frame rather than the space-fixed frame, defining the body-fixed
Euler angles $\zeta^{\rm bf}=(\alpha^{\rm bf},\beta^{\rm bf},\gamma^{\rm
bf})$. The rotation matrix that expresses the monomer frame axes in
body-fixed coordinates is given by $\boldsymbol{\mathcal{R}}^{\rm
bf}_{\rm mf}=\boldsymbol{\mathcal{R}}_{z}(\alpha^{\rm
bf})\boldsymbol{\mathcal{R}}_{y}(\beta^{\rm
bf})\boldsymbol{\mathcal{R}}_{z}(\gamma^{\rm bf})$. When the hydrogen
atoms are permuted, the body-fixed frame is unchanged, and as a result
the body-fixed Euler angles transform in precisely the same way as the
space-fixed ones. However, when $\hat{E}^{*}$ is applied, not only the
monomer frame axes in space-fixed coordinates change, but also the dimer
frame axes. As a result, the body-fixed Euler angles transform somewhat
differently than the space-fixed angles, as seen in Table \ref{tabcoor}.

Having determined the effect of the various symmetry operations on the
coordinates that describe the NH${}_3$-He complex, we can also find out
the corresponding effect on the angular basis functions of
Eqs.~(\ref{eq:bfbas}) and (\ref{eq:sfbas}) by using the transformation
properties of the Wigner $d$-functions. Moreover, we have that
$\hat{E}^*\phi^{\pm}_{v}(\rho)= \pm \phi^{\pm}_{v}(\rho)$. The effect of
the various symmetry operations on the angular basis functions is shown
in Table \ref{tabbas}.

As a result, we are now able to construct the symmetry adapted
bases sets for both the body-fixed and the space-fixed case. To this
end, it is most convenient to start by discussing the symmetry group
$C_{3v}({\rm M})$ with irreps $A_1$, $A_2$ and $E$. The $C_{3v}({\rm
M})$ group is generated by the operations $\hat{E}$, $(123)$ and
$(23)^*$. Using this group implies that we consider the ammonia molecule
as a rigid rotor without umbrella motion. Then, for the body-fixed case
with $k = K= 0$, we conclude from Table \ref{tabbas} that the state $| j
0 0 J M \rangle $ is of $A_1$ symmetry when $J$ is even, while it is of
$A_2$ symmetry when $J$ is odd. When either $k$ or $K$ is nonzero, we
have for $k=0$ (mod 3) that $(| j k K J M \rangle + (-1)^{J+k}| j {-k} {-K}
J M \rangle)/2^{1/2}$ is of $A_1$ symmetry, while $(| j k K J M \rangle
- (-1)^{J+k}| j {-k} {-K} J M \rangle)/2^{1/2}$ is of $A_2$ symmetry.
Finally, we have for $k \ne 0$ (mod 3), that two-dimensional $E$ irreps are spanned by the states $(| j k K J M \rangle,| j {-k} {-K} J
M \rangle)$.

For the space-fixed case and considering rigid ammonia, we conclude from
Table \ref{tabbas} that the state $| j 0 L J M \rangle $ is of $A_1$
symmetry when $j+L$ is even, while it is of $A_2$ symmetry when $j+L$ is
odd. When $k$ is nonzero and a multiple of 3, we have that $(| j k L J M
\rangle + (-1)^{j+L+k}| j {-k} L J M \rangle)/2^{1/2}$ is of $A_1$
symmetry, while $(| j k L J M \rangle - (-1)^{j+L+k}| j {-k} L J M
\rangle)/2^{1/2}$ is of $A_2$ symmetry. For $k \ne 0$
(mod 3), we have that two-dimensional $E$ irreps are spanned by the
states $\{| j k L J M \rangle,| j {-k} L J M \rangle \}$.

Finally, the complete symmetry adapted basis is obtained by considering
the full $D_{3h}({\rm M})$ symmetry group of the nonrigid ammonia
molecule. The functions adapted to the irreps of $D_{3h}({\rm M})$ are
obtained from those adapted to the $C_{3v}({\rm M})$ irreps by using
\begin{eqnarray}
&& (\hat{E}+\hat{E}^*)|A_1\rangle=|A_1'\rangle, \quad (\hat{E}-\hat{E}^*)|A_1\rangle=|A_2''\rangle, \nonumber \\
&& (\hat{E}+\hat{E}^*)|A_2\rangle=|A_2'\rangle,\nonumber \quad (\hat{E}-\hat{E}^*)|A_2\rangle=|A_1''\rangle, \\
&& (\hat{E}+\hat{E}^*)|E\rangle=|E' \rangle, \quad (\hat{E}-\hat{E}^*)|E\rangle=|E''\rangle.
\end{eqnarray}
Further we note that the $v^+$ umbrella functions belong to the
$(\hat{E}+\hat{E}^*)$ projection, while the $v^-$ umbrella functions belong to the $(\hat{E}-\hat{E}^*)$ projection.

\section*{Acknowledgements}

We thank Liesbeth Janssen for helpful discussions. We thank Hans-Joachim Werner for insightful correspondence on the long-range behavior of the CCSD(T)-F12 method. We thank Jeremy Hutson for informing us about the way in which the ammonia inversion model is implemented in {\sc molscat}. Koos Gubbels acknowledges support by the European Community's Seventh Framework Program ERC-2009-AdG under grant agreement 247142-MolChip. Sebastiaan Y.T. van de Meerakker acknowledges financial support from Netherlands Organisation for Scientific Research (NWO) via a VIDI grant. Ad van der Avoird thanks the Alexander von Humboldt
foundation for a Humboldt Research Award.


\begin{thebibliography}{57}
\expandafter\ifx\csname natexlab\endcsname\relax\def\natexlab#1{#1}\fi
\expandafter\ifx\csname bibnamefont\endcsname\relax
  \def\bibnamefont#1{#1}\fi
\expandafter\ifx\csname bibfnamefont\endcsname\relax
  \def\bibfnamefont#1{#1}\fi
\expandafter\ifx\csname citenamefont\endcsname\relax
  \def\citenamefont#1{#1}\fi
\expandafter\ifx\csname url\endcsname\relax
  \def\url#1{\texttt{#1}}\fi
\expandafter\ifx\csname urlprefix\endcsname\relax\def\urlprefix{URL }\fi
\providecommand{\bibinfo}[2]{#2}
\providecommand{\eprint}[2][]{\url{#2}}

\bibitem[{\citenamefont{Inguscio et~al.}(2008)\citenamefont{Inguscio, Ketterle,
  and Salomon}}]{inguscio:08}
\bibinfo{editor}{\bibfnamefont{M.}~\bibnamefont{Inguscio}},
  \bibinfo{editor}{\bibfnamefont{W.}~\bibnamefont{Ketterle}}, \bibnamefont{and}
  \bibinfo{editor}{\bibfnamefont{C.}~\bibnamefont{Salomon}}, eds.,
  \emph{\bibinfo{title}{Ultra-cold Fermi Gases}} (\bibinfo{publisher}{IOS
  Press}, \bibinfo{address}{Amsterdam}, \bibinfo{year}{2008}),
  \bibinfo{note}{proceedings of the International School of Physics Enrico
  Fermi, Course CLXIV, Varenna, 20-30 June 2006}.

\bibitem[{\citenamefont{Stoof et~al.}(2009)\citenamefont{Stoof, Gubbels, and
  Dickerscheid}}]{stoof:09}
\bibinfo{author}{\bibfnamefont{H.~T.~C.} \bibnamefont{Stoof}},
  \bibinfo{author}{\bibfnamefont{K.~B.} \bibnamefont{Gubbels}},
  \bibnamefont{and} \bibinfo{author}{\bibfnamefont{D.~B.~M.}
  \bibnamefont{Dickerscheid}}, \emph{\bibinfo{title}{Ultracold Quantum Fields}}
  (\bibinfo{publisher}{Springer}, \bibinfo{address}{Dordrecht},
  \bibinfo{year}{2009}).

\bibitem[{\citenamefont{Schutte et~al.}(1975)\citenamefont{Schutte, Bassi,
  Tommasini, and Scoles}}]{schutte:75a}
\bibinfo{author}{\bibfnamefont{A.}~\bibnamefont{Schutte}},
  \bibinfo{author}{\bibfnamefont{D.}~\bibnamefont{Bassi}},
  \bibinfo{author}{\bibfnamefont{F.}~\bibnamefont{Tommasini}},
  \bibnamefont{and} \bibinfo{author}{\bibfnamefont{G.}~\bibnamefont{Scoles}},
  \bibinfo{journal}{Phys. Rev. Lett.} \textbf{\bibinfo{volume}{29}},
  \bibinfo{pages}{979} (\bibinfo{year}{1975}).

\bibitem[{\citenamefont{Toennies et~al.}(1979)\citenamefont{Toennies, Welz, and
  Wolf}}]{toennies:79}
\bibinfo{author}{\bibfnamefont{J.~P.} \bibnamefont{Toennies}},
  \bibinfo{author}{\bibfnamefont{W.}~\bibnamefont{Welz}}, \bibnamefont{and}
  \bibinfo{author}{\bibfnamefont{G.}~\bibnamefont{Wolf}}, \bibinfo{journal}{J.
  Chem. Phys.} \textbf{\bibinfo{volume}{71}}, \bibinfo{pages}{614}
  (\bibinfo{year}{1979}).

\bibitem[{\citenamefont{Qiu et~al.}(2006)\citenamefont{Qiu, Ren, Che, Dai,
  Harich, Wang, Yang, Xu, Xie, Gustafsson et~al.}}]{qiu:06}
\bibinfo{author}{\bibfnamefont{M.~H.} \bibnamefont{Qiu}},
  \bibinfo{author}{\bibfnamefont{Z.~F.} \bibnamefont{Ren}},
  \bibinfo{author}{\bibfnamefont{L.}~\bibnamefont{Che}},
  \bibinfo{author}{\bibfnamefont{D.~X.} \bibnamefont{Dai}},
  \bibinfo{author}{\bibfnamefont{S.~A.} \bibnamefont{Harich}},
  \bibinfo{author}{\bibfnamefont{X.~Y.} \bibnamefont{Wang}},
  \bibinfo{author}{\bibfnamefont{X.~M.} \bibnamefont{Yang}},
  \bibinfo{author}{\bibfnamefont{C.~X.} \bibnamefont{Xu}},
  \bibinfo{author}{\bibfnamefont{D.~Q.} \bibnamefont{Xie}},
  \bibinfo{author}{\bibfnamefont{M.}~\bibnamefont{Gustafsson}},
  \bibnamefont{et~al.}, \bibinfo{journal}{Science}
  \textbf{\bibinfo{volume}{311}}, \bibinfo{pages}{1440} (\bibinfo{year}{2006}).

\bibitem[{\citenamefont{Dong et~al.}(2010)\citenamefont{Dong, Xiao, Wang, Dai,
  Yang, and Zhang}}]{Dong:Science327:1501}
\bibinfo{author}{\bibfnamefont{W.}~\bibnamefont{Dong}},
  \bibinfo{author}{\bibfnamefont{C.}~\bibnamefont{Xiao}},
  \bibinfo{author}{\bibfnamefont{T.}~\bibnamefont{Wang}},
  \bibinfo{author}{\bibfnamefont{D.~X.} \bibnamefont{Dai}},
  \bibinfo{author}{\bibfnamefont{X.~M.} \bibnamefont{Yang}}, \bibnamefont{and}
  \bibinfo{author}{\bibfnamefont{D.~H.} \bibnamefont{Zhang}},
  \bibinfo{journal}{Science} \textbf{\bibinfo{volume}{327}},
  \bibinfo{pages}{1501} (\bibinfo{year}{2010}).

\bibitem[{\citenamefont{Skodje et~al.}(2000{\natexlab{a}})\citenamefont{Skodje,
  Skouteris, Manolopoulos, Lee, Dong, and Liu}}]{skodje:00a}
\bibinfo{author}{\bibfnamefont{R.~T.} \bibnamefont{Skodje}},
  \bibinfo{author}{\bibfnamefont{D.}~\bibnamefont{Skouteris}},
  \bibinfo{author}{\bibfnamefont{D.~E.} \bibnamefont{Manolopoulos}},
  \bibinfo{author}{\bibfnamefont{S.-H.} \bibnamefont{Lee}},
  \bibinfo{author}{\bibfnamefont{F.}~\bibnamefont{Dong}}, \bibnamefont{and}
  \bibinfo{author}{\bibfnamefont{K.}~\bibnamefont{Liu}},
  \bibinfo{journal}{Phys. Rev. Lett.} \textbf{\bibinfo{volume}{85}},
  \bibinfo{pages}{1206} (\bibinfo{year}{2000}{\natexlab{a}}).

\bibitem[{\citenamefont{Skodje et~al.}(2000{\natexlab{b}})\citenamefont{Skodje,
  Skouteris, Manolopoulos, Lee, Dong, and Liu}}]{skodje:00b}
\bibinfo{author}{\bibfnamefont{R.~T.} \bibnamefont{Skodje}},
  \bibinfo{author}{\bibfnamefont{D.}~\bibnamefont{Skouteris}},
  \bibinfo{author}{\bibfnamefont{D.~E.} \bibnamefont{Manolopoulos}},
  \bibinfo{author}{\bibfnamefont{S.-H.} \bibnamefont{Lee}},
  \bibinfo{author}{\bibfnamefont{F.}~\bibnamefont{Dong}}, \bibnamefont{and}
  \bibinfo{author}{\bibfnamefont{K.}~\bibnamefont{Liu}}, \bibinfo{journal}{J.
  Chem. Phys.} \textbf{\bibinfo{volume}{112}}, \bibinfo{pages}{4536}
  (\bibinfo{year}{2000}{\natexlab{b}}).

\bibitem[{\citenamefont{Gilijamse et~al.}(2006)\citenamefont{Gilijamse,
  Hoekstra, van~de Meerakker, Groenenboom, and Meijer}}]{gilijamse:06}
\bibinfo{author}{\bibfnamefont{J.~J.} \bibnamefont{Gilijamse}},
  \bibinfo{author}{\bibfnamefont{S.}~\bibnamefont{Hoekstra}},
  \bibinfo{author}{\bibfnamefont{S.~Y.~T.} \bibnamefont{van~de Meerakker}},
  \bibinfo{author}{\bibfnamefont{G.~C.} \bibnamefont{Groenenboom}},
  \bibnamefont{and} \bibinfo{author}{\bibfnamefont{G.}~\bibnamefont{Meijer}},
  \bibinfo{journal}{Science} \textbf{\bibinfo{volume}{313}},
  \bibinfo{pages}{1617} (\bibinfo{year}{2006}).

\bibitem[{\citenamefont{Bethlem et~al.}(1999)\citenamefont{Bethlem, Berden, and
  Meijer}}]{bethlem:99}
\bibinfo{author}{\bibfnamefont{H.~L.} \bibnamefont{Bethlem}},
  \bibinfo{author}{\bibfnamefont{G.}~\bibnamefont{Berden}}, \bibnamefont{and}
  \bibinfo{author}{\bibfnamefont{G.}~\bibnamefont{Meijer}},
  \bibinfo{journal}{Phys. Rev. Lett.} \textbf{\bibinfo{volume}{83}},
  \bibinfo{pages}{1558} (\bibinfo{year}{1999}).

\bibitem[{\citenamefont{Scharfenberg et~al.}(2010)\citenamefont{Scharfenberg,
  K{\l}os, Dagdigian, Alexander, Meijer, and van~de
  Meerakker}}]{scharfenberg:10}
\bibinfo{author}{\bibfnamefont{L.}~\bibnamefont{Scharfenberg}},
  \bibinfo{author}{\bibfnamefont{J.}~\bibnamefont{K{\l}os}},
  \bibinfo{author}{\bibfnamefont{P.~J.} \bibnamefont{Dagdigian}},
  \bibinfo{author}{\bibfnamefont{M.~H.} \bibnamefont{Alexander}},
  \bibinfo{author}{\bibfnamefont{G.}~\bibnamefont{Meijer}}, \bibnamefont{and}
  \bibinfo{author}{\bibfnamefont{S.~Y.~T.} \bibnamefont{van~de Meerakker}},
  \bibinfo{journal}{Phys. Chem. Chem. Phys.} \textbf{\bibinfo{volume}{12}},
  \bibinfo{pages}{10660} (\bibinfo{year}{2010}).

\bibitem[{\citenamefont{Kirste et~al.}(2010)\citenamefont{Kirste, Scharfenberg,
  Klos, Lique, Alexander, Meijer, and van~de Meerakker}}]{kirste:10}
\bibinfo{author}{\bibfnamefont{M.}~\bibnamefont{Kirste}},
  \bibinfo{author}{\bibfnamefont{L.}~\bibnamefont{Scharfenberg}},
  \bibinfo{author}{\bibfnamefont{J.}~\bibnamefont{Klos}},
  \bibinfo{author}{\bibfnamefont{F.}~\bibnamefont{Lique}},
  \bibinfo{author}{\bibfnamefont{M.~H.} \bibnamefont{Alexander}},
  \bibinfo{author}{\bibfnamefont{G.}~\bibnamefont{Meijer}}, \bibnamefont{and}
  \bibinfo{author}{\bibfnamefont{S.~Y.~T.} \bibnamefont{van~de Meerakker}},
  \bibinfo{journal}{Phys. Rev. A} \textbf{\bibinfo{volume}{82}},
  \bibinfo{pages}{042717} (\bibinfo{year}{2010}).

\bibitem[{\citenamefont{Scharfenberg
  et~al.}(2011{\natexlab{a}})\citenamefont{Scharfenberg, Gubbels, Kirste,
  Groenenboom, van~der Avoird, Meijer, and van~de Meerakker}}]{scharfenberg:11}
\bibinfo{author}{\bibfnamefont{L.}~\bibnamefont{Scharfenberg}},
  \bibinfo{author}{\bibfnamefont{K.~B.} \bibnamefont{Gubbels}},
  \bibinfo{author}{\bibfnamefont{M.}~\bibnamefont{Kirste}},
  \bibinfo{author}{\bibfnamefont{G.~C.} \bibnamefont{Groenenboom}},
  \bibinfo{author}{\bibfnamefont{A.}~\bibnamefont{van~der Avoird}},
  \bibinfo{author}{\bibfnamefont{G.}~\bibnamefont{Meijer}}, \bibnamefont{and}
  \bibinfo{author}{\bibfnamefont{S.~Y.~T.} \bibnamefont{van~de Meerakker}},
  \bibinfo{journal}{Eur. Phys. J. D} p. \bibinfo{pages}{accepted}
  (\bibinfo{year}{2011}{\natexlab{a}}), \eprint{arXiv:1101.0948}.

\bibitem[{\citenamefont{{van der Avoird} et~al.}(1994)\citenamefont{{van der
  Avoird}, Wormer, and Moszynski}}]{avoird:94}
\bibinfo{author}{\bibfnamefont{A.}~\bibnamefont{{van der Avoird}}},
  \bibinfo{author}{\bibfnamefont{P.~E.~S.} \bibnamefont{Wormer}},
  \bibnamefont{and}
  \bibinfo{author}{\bibfnamefont{R.}~\bibnamefont{Moszynski}},
  \bibinfo{journal}{Chem. Rev.} \textbf{\bibinfo{volume}{94}},
  \bibinfo{pages}{1931} (\bibinfo{year}{1994}).

\bibitem[{\citenamefont{Wormer and van~der Avoird}(2000)}]{wormer:00a}
\bibinfo{author}{\bibfnamefont{P.~E.~S.} \bibnamefont{Wormer}}
  \bibnamefont{and} \bibinfo{author}{\bibfnamefont{A.}~\bibnamefont{van~der
  Avoird}}, \bibinfo{journal}{Chem. Rev.} \textbf{\bibinfo{volume}{100}},
  \bibinfo{pages}{4109} (\bibinfo{year}{2000}).

\bibitem[{\citenamefont{Slipchenko and Vilesov}(2005)}]{slipchenko:05}
\bibinfo{author}{\bibfnamefont{M.}~\bibnamefont{Slipchenko}} \bibnamefont{and}
  \bibinfo{author}{\bibfnamefont{A.}~\bibnamefont{Vilesov}},
  \bibinfo{journal}{Chem. Phys. Lett.} \textbf{\bibinfo{volume}{412}},
  \bibinfo{pages}{176} (\bibinfo{year}{2005}).

\bibitem[{\citenamefont{Cheung et~al.}(1968)\citenamefont{Cheung, Rank, Townes,
  Thornton, and Welch}}]{cheung:68}
\bibinfo{author}{\bibfnamefont{A.~C.} \bibnamefont{Cheung}},
  \bibinfo{author}{\bibfnamefont{D.~M.} \bibnamefont{Rank}},
  \bibinfo{author}{\bibfnamefont{C.~H.} \bibnamefont{Townes}},
  \bibinfo{author}{\bibfnamefont{D.~D.} \bibnamefont{Thornton}},
  \bibnamefont{and} \bibinfo{author}{\bibfnamefont{W.~J.} \bibnamefont{Welch}},
  \bibinfo{journal}{Phys. Rev. Lett.} \textbf{\bibinfo{volume}{21}},
  \bibinfo{pages}{1701} (\bibinfo{year}{1968}).

\bibitem[{\citenamefont{Oka}(1968)}]{oka:68}
\bibinfo{author}{\bibfnamefont{T.}~\bibnamefont{Oka}}, \bibinfo{journal}{J.
  Chem. Phys.} \textbf{\bibinfo{volume}{49}}, \bibinfo{pages}{3135}
  (\bibinfo{year}{1968}).

\bibitem[{\citenamefont{Meyer et~al.}(1986)\citenamefont{Meyer, Buck, Schinke,
  and Diercksen}}]{meyer:86}
\bibinfo{author}{\bibfnamefont{H.}~\bibnamefont{Meyer}},
  \bibinfo{author}{\bibfnamefont{U.}~\bibnamefont{Buck}},
  \bibinfo{author}{\bibfnamefont{R.}~\bibnamefont{Schinke}}, \bibnamefont{and}
  \bibinfo{author}{\bibfnamefont{G.}~\bibnamefont{Diercksen}},
  \bibinfo{journal}{J. Chem. Phys.} \textbf{\bibinfo{volume}{86}},
  \bibinfo{pages}{4976} (\bibinfo{year}{1986}).

\bibitem[{\citenamefont{Seelemann et~al.}(1988)\citenamefont{Seelemann,
  Andresen, Schleipen, Beyer, and ter Meulen}}]{seelemann:88}
\bibinfo{author}{\bibfnamefont{T.}~\bibnamefont{Seelemann}},
  \bibinfo{author}{\bibfnamefont{P.}~\bibnamefont{Andresen}},
  \bibinfo{author}{\bibfnamefont{J.}~\bibnamefont{Schleipen}},
  \bibinfo{author}{\bibfnamefont{B.}~\bibnamefont{Beyer}}, \bibnamefont{and}
  \bibinfo{author}{\bibfnamefont{J.~J.} \bibnamefont{ter Meulen}},
  \bibinfo{journal}{Chem. Phys.} \textbf{\bibinfo{volume}{126}},
  \bibinfo{pages}{27} (\bibinfo{year}{1988}).

\bibitem[{\citenamefont{Schleipen and ter Meulen}(1991)}]{schleipen:91}
\bibinfo{author}{\bibfnamefont{J.}~\bibnamefont{Schleipen}} \bibnamefont{and}
  \bibinfo{author}{\bibfnamefont{J.~J.} \bibnamefont{ter Meulen}},
  \bibinfo{journal}{Chem. Phys.} \textbf{\bibinfo{volume}{156}},
  \bibinfo{pages}{479} (\bibinfo{year}{1991}).

\bibitem[{\citenamefont{Meyer}(1995)}]{meyer:95}
\bibinfo{author}{\bibfnamefont{H.}~\bibnamefont{Meyer}}, \bibinfo{journal}{J.
  Chem. Phys.} \textbf{\bibinfo{volume}{99}}, \bibinfo{pages}{1101}
  (\bibinfo{year}{1995}).

\bibitem[{\citenamefont{Green}(1976)}]{green:76}
\bibinfo{author}{\bibfnamefont{S.}~\bibnamefont{Green}}, \bibinfo{journal}{J.
  Chem. Phys.} \textbf{\bibinfo{volume}{64}}, \bibinfo{pages}{3463}
  (\bibinfo{year}{1976}).

\bibitem[{\citenamefont{Billing et~al.}(1985)\citenamefont{Billing, Poulsen,
  and Diercksen}}]{billing:85}
\bibinfo{author}{\bibfnamefont{G.~D.} \bibnamefont{Billing}},
  \bibinfo{author}{\bibfnamefont{L.~L.} \bibnamefont{Poulsen}},
  \bibnamefont{and} \bibinfo{author}{\bibfnamefont{G.~H.~F.}
  \bibnamefont{Diercksen}}, \bibinfo{journal}{Chem. Phys.}
  \textbf{\bibinfo{volume}{98}}, \bibinfo{pages}{397} (\bibinfo{year}{1985}).

\bibitem[{\citenamefont{Chen and Zhang}(1997)}]{chen:97}
\bibinfo{author}{\bibfnamefont{J.}~\bibnamefont{Chen}} \bibnamefont{and}
  \bibinfo{author}{\bibfnamefont{Y.}~\bibnamefont{Zhang}}, \bibinfo{journal}{J.
  Phys. B: At. Mol. Opt. Phys.} \textbf{\bibinfo{volume}{30}},
  \bibinfo{pages}{347} (\bibinfo{year}{1997}).

\bibitem[{\citenamefont{Wang}(2003)}]{wang:03}
\bibinfo{author}{\bibfnamefont{W.~F.} \bibnamefont{Wang}},
  \bibinfo{journal}{Chem. Phys.} \textbf{\bibinfo{volume}{288}},
  \bibinfo{pages}{23} (\bibinfo{year}{2003}).

\bibitem[{\citenamefont{Machin and Roueff}(2005)}]{machin:05}
\bibinfo{author}{\bibfnamefont{U.}~\bibnamefont{Machin}} \bibnamefont{and}
  \bibinfo{author}{\bibfnamefont{E.}~\bibnamefont{Roueff}},
  \bibinfo{journal}{J. Phys. B: At. Mol. Opt. Phys.}
  \textbf{\bibinfo{volume}{30}}, \bibinfo{pages}{1519} (\bibinfo{year}{2005}).

\bibitem[{\citenamefont{Yang and Stancil}(2008)}]{yang:08}
\bibinfo{author}{\bibfnamefont{B.}~\bibnamefont{Yang}} \bibnamefont{and}
  \bibinfo{author}{\bibfnamefont{P.}~\bibnamefont{Stancil}},
  \bibinfo{journal}{Eur. Phys. J. D} \textbf{\bibinfo{volume}{47}},
  \bibinfo{pages}{351} (\bibinfo{year}{2008}).

\bibitem[{\citenamefont{Hodges and Wheatley}(2001)}]{hodges:01}
\bibinfo{author}{\bibfnamefont{M.~P.} \bibnamefont{Hodges}} \bibnamefont{and}
  \bibinfo{author}{\bibfnamefont{R.~J.} \bibnamefont{Wheatley}},
  \bibinfo{journal}{J. Chem. Phys.} \textbf{\bibinfo{volume}{114}},
  \bibinfo{pages}{8836} (\bibinfo{year}{2001}).

\bibitem[{\citenamefont{{van Bladel} et~al.}(1991)\citenamefont{{van Bladel},
  {van der Avoird}, and Wormer}}]{bladel:91}
\bibinfo{author}{\bibfnamefont{J.~W.~I.} \bibnamefont{{van Bladel}}},
  \bibinfo{author}{\bibfnamefont{A.}~\bibnamefont{{van der Avoird}}},
  \bibnamefont{and} \bibinfo{author}{\bibfnamefont{P.~E.~S.}
  \bibnamefont{Wormer}}, \bibinfo{journal}{J. Phys. Chem.}
  \textbf{\bibinfo{volume}{95}}, \bibinfo{pages}{5414} (\bibinfo{year}{1991}).

\bibitem[{\citenamefont{{van Bladel} et~al.}(1992)\citenamefont{{van Bladel},
  {van der Avoird}, and Wormer}}]{bladel:92a}
\bibinfo{author}{\bibfnamefont{J.~W.~I.} \bibnamefont{{van Bladel}}},
  \bibinfo{author}{\bibfnamefont{A.}~\bibnamefont{{van der Avoird}}},
  \bibnamefont{and} \bibinfo{author}{\bibfnamefont{P.~E.~S.}
  \bibnamefont{Wormer}}, \bibinfo{journal}{Chem. Phys.}
  \textbf{\bibinfo{volume}{165}}, \bibinfo{pages}{47} (\bibinfo{year}{1992}).

\bibitem[{\citenamefont{Huang et~al.}(2008)\citenamefont{Huang, Schwenke, and
  Lee}}]{huang:08}
\bibinfo{author}{\bibfnamefont{X.}~\bibnamefont{Huang}},
  \bibinfo{author}{\bibfnamefont{D.~W.} \bibnamefont{Schwenke}},
  \bibnamefont{and} \bibinfo{author}{\bibfnamefont{T.~J.} \bibnamefont{Lee}},
  \bibinfo{journal}{J. Chem. Phys.} \textbf{\bibinfo{volume}{129}},
  \bibinfo{pages}{214304} (\bibinfo{year}{2008}).

\bibitem[{\citenamefont{Townes and Schawlow}(1975)}]{townes:75}
\bibinfo{author}{\bibfnamefont{C.~H.} \bibnamefont{Townes}} \bibnamefont{and}
  \bibinfo{author}{\bibfnamefont{A.~L.} \bibnamefont{Schawlow}},
  \emph{\bibinfo{title}{Microwave Spectroscopy}} (\bibinfo{publisher}{Dover},
  \bibinfo{address}{New York}, \bibinfo{year}{1975}).

\bibitem[{\citenamefont{Bunker and Jensen}(1998)}]{bunker:98}
\bibinfo{author}{\bibfnamefont{P.~R.} \bibnamefont{Bunker}} \bibnamefont{and}
  \bibinfo{author}{\bibfnamefont{P.}~\bibnamefont{Jensen}},
  \emph{\bibinfo{title}{Molecular Symmetry and Spectroscopy}}
  (\bibinfo{publisher}{NRC Research Press}, \bibinfo{address}{Ottawa},
  \bibinfo{year}{1998}), \bibinfo{edition}{2nd} ed.

\bibitem[{\citenamefont{Johnson}(1973)}]{johnson:73}
\bibinfo{author}{\bibfnamefont{B.~R.} \bibnamefont{Johnson}},
  \bibinfo{journal}{J. Comp. Phys.} \textbf{\bibinfo{volume}{13}},
  \bibinfo{pages}{445} (\bibinfo{year}{1973}).

\bibitem[{\citenamefont{Child}(1974)}]{child:74}
\bibinfo{author}{\bibfnamefont{M.~S.} \bibnamefont{Child}},
  \emph{\bibinfo{title}{Molecular Collision Theory}}
  (\bibinfo{publisher}{Academic Press}, \bibinfo{address}{London},
  \bibinfo{year}{1974}).

\bibitem[{\citenamefont{Werner et~al.}()\citenamefont{Werner, Knowles, and {\em
  et al.}}}]{molpro:09}
\bibinfo{author}{\bibfnamefont{H.-J.} \bibnamefont{Werner}},
  \bibinfo{author}{\bibfnamefont{P.~J.} \bibnamefont{Knowles}},
  \bibnamefont{and} \bibinfo{author}{\bibnamefont{{\em et al.}}},
  \emph{\bibinfo{title}{{\sc molpro}: a package of ab initio programs, version
  2009.1}}, \urlprefix\url{http://www.molpro.net}.

\bibitem[{\citenamefont{Boys and Bernardi}(1970)}]{boys:70}
\bibinfo{author}{\bibfnamefont{S.~F.} \bibnamefont{Boys}} \bibnamefont{and}
  \bibinfo{author}{\bibfnamefont{F.}~\bibnamefont{Bernardi}},
  \bibinfo{journal}{Mol. Phys.} \textbf{\bibinfo{volume}{19}},
  \bibinfo{pages}{553} (\bibinfo{year}{1970}).

\bibitem[{\citenamefont{Adler et~al.}(2007)\citenamefont{Adler, Knizia, and
  H.-J.Werner}}]{adler:07}
\bibinfo{author}{\bibfnamefont{T.~B.} \bibnamefont{Adler}},
  \bibinfo{author}{\bibfnamefont{G.}~\bibnamefont{Knizia}}, \bibnamefont{and}
  \bibinfo{author}{\bibnamefont{H.-J.Werner}}, \bibinfo{journal}{J. Chem.
  Phys.} \textbf{\bibinfo{volume}{127}}, \bibinfo{pages}{221106}
  (\bibinfo{year}{2007}).

\bibitem[{\citenamefont{Koch et~al.}(1998)\citenamefont{Koch, Fern{\'a}ndez,
  and Christiansen}}]{koch:98}
\bibinfo{author}{\bibfnamefont{H.}~\bibnamefont{Koch}},
  \bibinfo{author}{\bibfnamefont{B.}~\bibnamefont{Fern{\'a}ndez}},
  \bibnamefont{and}
  \bibinfo{author}{\bibfnamefont{O.}~\bibnamefont{Christiansen}},
  \bibinfo{journal}{J. Chem. Phys.} \textbf{\bibinfo{volume}{108}},
  \bibinfo{pages}{2784} (\bibinfo{year}{1998}).

\bibitem[{\citenamefont{van~der Avoird et~al.}(1980)\citenamefont{van~der
  Avoird, Wormer, Mulder, and Berns}}]{avoird:80}
\bibinfo{author}{\bibfnamefont{A.}~\bibnamefont{van~der Avoird}},
  \bibinfo{author}{\bibfnamefont{P.~E.~S.} \bibnamefont{Wormer}},
  \bibinfo{author}{\bibfnamefont{F.}~\bibnamefont{Mulder}}, \bibnamefont{and}
  \bibinfo{author}{\bibfnamefont{R.~M.} \bibnamefont{Berns}},
  \bibinfo{journal}{Top. Curr. Chem.} \textbf{\bibinfo{volume}{93}},
  \bibinfo{pages}{1} (\bibinfo{year}{1980}).

\bibitem[{\citenamefont{Tang and Toennies}(1984)}]{tang:84}
\bibinfo{author}{\bibfnamefont{K.~T.} \bibnamefont{Tang}} \bibnamefont{and}
  \bibinfo{author}{\bibfnamefont{J.~P.} \bibnamefont{Toennies}},
  \bibinfo{journal}{J. Chem. Phys.} \textbf{\bibinfo{volume}{80}},
  \bibinfo{pages}{3726} (\bibinfo{year}{1984}).

\bibitem[{\citenamefont{Ho and Rabitz}(1996)}]{ho:96}
\bibinfo{author}{\bibfnamefont{T.-S.} \bibnamefont{Ho}} \bibnamefont{and}
  \bibinfo{author}{\bibfnamefont{H.}~\bibnamefont{Rabitz}},
  \bibinfo{journal}{J. Chem. Phys.} \textbf{\bibinfo{volume}{104}},
  \bibinfo{pages}{2584} (\bibinfo{year}{1996}).

\bibitem[{epa()}]{epaps:pot}
\bibinfo{note}{See supplementary material at [URL will be inserted by AIP] for
  the fit of the potential energy surface and for the {\it ab initio} data
  points.}

\bibitem[{\citenamefont{Groenenboom and Colbert}(1993)}]{groenenboom:93}
\bibinfo{author}{\bibfnamefont{G.~C.} \bibnamefont{Groenenboom}}
  \bibnamefont{and} \bibinfo{author}{\bibfnamefont{D.~T.}
  \bibnamefont{Colbert}}, \bibinfo{journal}{J. Chem. Phys.}
  \textbf{\bibinfo{volume}{99}}, \bibinfo{pages}{9681} (\bibinfo{year}{1993}).

\bibitem[{\citenamefont{Davis and Boggs}(1978)}]{davis:78}
\bibinfo{author}{\bibfnamefont{S.~L.} \bibnamefont{Davis}} \bibnamefont{and}
  \bibinfo{author}{\bibfnamefont{J.~E.} \bibnamefont{Boggs}},
  \bibinfo{journal}{J. Chem. Phys.} \textbf{\bibinfo{volume}{69}},
  \bibinfo{pages}{2355} (\bibinfo{year}{1978}).

\bibitem[{\citenamefont{Green}(1980)}]{green:80}
\bibinfo{author}{\bibfnamefont{S.}~\bibnamefont{Green}}, \bibinfo{journal}{J.
  Chem. Phys.} \textbf{\bibinfo{volume}{73}}, \bibinfo{pages}{2740}
  (\bibinfo{year}{1980}).

\bibitem[{\citenamefont{Hutson and Green}()}]{molscat:94}
\bibinfo{author}{\bibfnamefont{J.~M.} \bibnamefont{Hutson}} \bibnamefont{and}
  \bibinfo{author}{\bibfnamefont{S.}~\bibnamefont{Green}}, \bibinfo{note}{{\sc
  molscat} computer code, version 14 (1994), distributed by Collaborative
  Computational Project No. 6 of the Engineering and Physical Sciences Research
  Council (UK)}.

\bibitem[{\citenamefont{{van der Sanden} et~al.}(1992)\citenamefont{{van der
  Sanden}, Wormer, {van der Avoird}, Schleipen, and {ter Meulen}}}]{sanden:93}
\bibinfo{author}{\bibfnamefont{G.~C.~M.} \bibnamefont{{van der Sanden}}},
  \bibinfo{author}{\bibfnamefont{P.~E.~S.} \bibnamefont{Wormer}},
  \bibinfo{author}{\bibfnamefont{A.}~\bibnamefont{{van der Avoird}}},
  \bibinfo{author}{\bibfnamefont{J.}~\bibnamefont{Schleipen}},
  \bibnamefont{and} \bibinfo{author}{\bibfnamefont{J.~J.} \bibnamefont{{ter
  Meulen}}}, \bibinfo{journal}{J. Chem. Phys.} \textbf{\bibinfo{volume}{97}},
  \bibinfo{pages}{6460} (\bibinfo{year}{1992}).

\bibitem[{\citenamefont{Wigner}(1948)}]{wigner:48}
\bibinfo{author}{\bibfnamefont{E.~P.} \bibnamefont{Wigner}},
  \bibinfo{journal}{Phys. Rev.} \textbf{\bibinfo{volume}{73}},
  \bibinfo{pages}{1002} (\bibinfo{year}{1948}).

\bibitem[{\citenamefont{Ashton et~al.}(1983)\citenamefont{Ashton, Child, and
  Hutson}}]{ashton:83}
\bibinfo{author}{\bibfnamefont{C.~J.} \bibnamefont{Ashton}},
  \bibinfo{author}{\bibfnamefont{M.~S.} \bibnamefont{Child}}, \bibnamefont{and}
  \bibinfo{author}{\bibfnamefont{J.~M.} \bibnamefont{Hutson}},
  \bibinfo{journal}{J. Chem. Phys.} \textbf{\bibinfo{volume}{78}},
  \bibinfo{pages}{4025} (\bibinfo{year}{1983}).

\bibitem[{\citenamefont{Millan et~al.}(1995)\citenamefont{Millan, Halberstadt,
  van~der Sanden, and {van der Avoird}}}]{millan:95}
\bibinfo{author}{\bibfnamefont{J.}~\bibnamefont{Millan}},
  \bibinfo{author}{\bibfnamefont{N.}~\bibnamefont{Halberstadt}},
  \bibinfo{author}{\bibfnamefont{G.~C.~M.} \bibnamefont{van~der Sanden}},
  \bibnamefont{and} \bibinfo{author}{\bibfnamefont{A.}~\bibnamefont{{van der
  Avoird}}}, \bibinfo{journal}{J. Chem. Phys.} \textbf{\bibinfo{volume}{103}},
  \bibinfo{pages}{4138} (\bibinfo{year}{1995}).

\bibitem[{\citenamefont{Bethlem et~al.}(2002)\citenamefont{Bethlem, Crompvoets,
  Jongma, van~de Meerakker, and Meijer}}]{Bethlem:PRA65:053416}
\bibinfo{author}{\bibfnamefont{H.~L.} \bibnamefont{Bethlem}},
  \bibinfo{author}{\bibfnamefont{F.~M.~H.} \bibnamefont{Crompvoets}},
  \bibinfo{author}{\bibfnamefont{R.~T.} \bibnamefont{Jongma}},
  \bibinfo{author}{\bibfnamefont{S.~Y.~T.} \bibnamefont{van~de Meerakker}},
  \bibnamefont{and} \bibinfo{author}{\bibfnamefont{G.}~\bibnamefont{Meijer}},
  \bibinfo{journal}{Phys. Rev. A} \textbf{\bibinfo{volume}{65}},
  \bibinfo{pages}{053416} (\bibinfo{year}{2002}).

\bibitem[{\citenamefont{Heiner et~al.}(2006)\citenamefont{Heiner, Bethlem, and
  Meijer}}]{Heiner:PCCP8:2666}
\bibinfo{author}{\bibfnamefont{C.~E.} \bibnamefont{Heiner}},
  \bibinfo{author}{\bibfnamefont{H.~L.} \bibnamefont{Bethlem}},
  \bibnamefont{and} \bibinfo{author}{\bibfnamefont{G.}~\bibnamefont{Meijer}},
  \bibinfo{journal}{Phys. Chem. Chem. Phys.} \textbf{\bibinfo{volume}{8}},
  \bibinfo{pages}{2666} (\bibinfo{year}{2006}).

\bibitem[{\citenamefont{Scharfenberg
  et~al.}(2011{\natexlab{b}})\citenamefont{Scharfenberg, van~de Meerakker, and
  Meijer}}]{Scharfenberg:PCCP13:8448}
\bibinfo{author}{\bibfnamefont{L.}~\bibnamefont{Scharfenberg}},
  \bibinfo{author}{\bibfnamefont{S.~Y.~T.} \bibnamefont{van~de Meerakker}},
  \bibnamefont{and} \bibinfo{author}{\bibfnamefont{G.}~\bibnamefont{Meijer}},
  \bibinfo{journal}{Phys. Chem. Chem. Phys.} \textbf{\bibinfo{volume}{13}},
  \bibinfo{pages}{8448} (\bibinfo{year}{2011}{\natexlab{b}}).

\bibitem[{\citenamefont{Scharfenberg et~al.}(2009)\citenamefont{Scharfenberg,
  Haak, Meijer, and van~de Meerakker}}]{Scharfenberg:PRA79:023410}
\bibinfo{author}{\bibfnamefont{L.}~\bibnamefont{Scharfenberg}},
  \bibinfo{author}{\bibfnamefont{H.}~\bibnamefont{Haak}},
  \bibinfo{author}{\bibfnamefont{G.}~\bibnamefont{Meijer}}, \bibnamefont{and}
  \bibinfo{author}{\bibfnamefont{S.~Y.~T.} \bibnamefont{van~de Meerakker}},
  \bibinfo{journal}{Phys. Rev. A} \textbf{\bibinfo{volume}{79}},
  \bibinfo{pages}{023410} (\bibinfo{year}{2009}).

\bibitem[{\citenamefont{Eppink and Parker}(1997)}]{eppink:97}
\bibinfo{author}{\bibfnamefont{A.~T. J.~B.} \bibnamefont{Eppink}}
  \bibnamefont{and} \bibinfo{author}{\bibfnamefont{D.~H.}
  \bibnamefont{Parker}}, \bibinfo{journal}{Rev. Sci. Instrum.}
  \textbf{\bibinfo{volume}{68}}, \bibinfo{pages}{3477} (\bibinfo{year}{1997}).

\end{thebibliography}
\end{document}